\documentclass[useAMS,usenatbib]{mn2e}

\usepackage{graphicx}
\usepackage{multirow}
\usepackage{amssymb}
\usepackage{hyperref}

\title[\od\/ in UGC12846]{The Type IIP SN 2007od in UGC 12846: from a bright maximum to dust formation in the nebular phase\thanks{Based on observations collected at the Italian 3.58m Telescopio Nazionale Galileo, the Liverpool Telescope and the North Optical Telescope (La Palma, Spain), the Calar Alto Observatory (Sierra de los Filabres, Spain), the orbital Telescope SWIFT and at the Italian Rapid Eye Movement Telescope (La Silla, Chile) and the Copernico and Galileo Galilei telescopes (Asiago, Italy).}}
\author[C.Inserra et al.]{C. Inserra$^{1}$$^{,}$$^{2}$$^{,}$$^{3}$\thanks{E-mail: cosimo.inserra@oact.inaf.it(CI)} , M. Turatto$^{2,4}$, A. Pastorello$^{5}$, S. Benetti$^{6}$, E. Cappellaro$^{6}$, M.L. Pumo$^{6}$,
\newauthor L.Zampieri$^{6}$, I. Agnoletto$^{6}$, F. Bufano$^{6}$,  M.T. Botticella$^{6}$, M. Della Valle$^{7}$, N. Elias Rosa$^{8}$, 
\newauthor T. Iijima$^{9}$, S. Spiro$^{10}$, S. Valenti$^{5}$. 
\\
\\
$^{1}$Dipartimento di Fisica ed Astronomia, Universit\'a di Catania, Sezione Astrofisica, Via S.Sofia 78, 95123, Catania, Italy\\
$^{2}$INAF Osservatorio Astrofisico di Catania, Via S.Sofia 78, 95123, Catania, Italy\\
$^{3}$Department of Physics and Astronomy, University of Oklahoma, Norman, OK 73019, USA\\
$^{4}$INAF Osservatorio Astronomico di Trieste, Via Tiepolo 11, 34143, Trieste, Italy\\
$^{5}$Astrophysics Research Centre, School of Mathematics and Physics, Queen's University Belfast, Belfast BT7 1NN, United Kingdom\\
$^{6}$INAF Osservatorio Astronomico di Padova, Vicolo dell'Osservatotio 5, 35122, Padova, Italy\\
$^{7}$INAF Osservatorio Astronomico di Capodimonte, Salita Moiariello 16, 80131 Napoli, Italy\\
$^{8}$Institut de Cincies de l'Espai (IEEC-CSIC), Campus UAB, 08193 Bellaterra, Spain\\
$^{9}$INAF Osservatorio Astronomico di Padova, Sezione di Asiago, Osservatorio Astrofisico, 36012 Asiago (Vi), Italy\\
$^{10}$INAF National Institute for Astrophysics, I-00136 Rome, Italy}

\def\kms{km\,s$^{-1}$}
\def\Ha{H$\alpha$}
\def\Hb{H$\beta$}
\def\Hg{H$\gamma$}
\def\Hd{H$\delta$}
\def\ni{$^{56}$Ni}
\def\co{$^{56}$Co}
\def\fe{$^{56}$Fe}
\def\a{SN~1987A}
\def\em{SN~1999em}
\def\et{SN~2004et}
\def\h{SN~1992H}
\def\od{SN~2007od}
\def\cs{SN~2005cs}

\def\s{SN~1998S}
\def\mcento{mag\,(100d)$^{-1}$}

\def\M{M$_{\odot}$}

\def\ebv{E(B--V)}

\begin{document}

\date{Received.....; accepted.........}

\pagerange{\pageref{firstpage}--\pageref{lastpage}} \pubyear{???}

\maketitle

\label{firstpage}

\begin{abstract}

  Ultraviolet (UV), optical, and near infrared (NIR) observations of
  the Type IIP supernova (SN) 2007od, covering from maximum light to
  late phases, allow detailed investigation of different physical
  phenomena in the expanding ejecta. These data turn this object into
  one of the most peculiar SNe IIP ever studied. The early light curve
  of \od\/ is similar to that of a bright IIP, with a short plateau, a
  bright peak (M$_V=-$18 mag), but a very faint late--time optical
  light curve. However, with the inclusion of mid infrared (MIR)
  observations during the radioactive tail, we derive an ejected mass of \ni\/ of
  M($^{56}$Ni)$\sim 2 \times 10^{-2}$ \M. By modeling the bolometric
  light curve, ejecta expansion velocities, and blackbody temperature,
  we estimate a total ejected mass of 5 -- 7.5 M$_{\odot}$ with a
  kinetic energy of at least 0.5 $\times$ 10$^{51}$ erg. The early
  spectra reveal a boxy \Ha\/ profile and high velocity
  features of the Balmer series that suggest the possible interaction of the ejecta with 
  a close circumstellar matter (CSM). 
  The interaction with the CSM and the
  presence of dust formed inside the ejecta are evident in the late-time spectra.
  The episodes of mass loss
  shortly before explosion, the bright plateau, the relatively small
  amount of \ni, and the faint [O I] emission observed in the nebular
  spectra are consistent with a super-asymptotic giant branch
  (super-AGB) progenitor (M$\sim$9.7 - 11 \M).

\end{abstract}


\section{Introduction}\label{sec:intro}

Type II supernovae (SNe II) are produced in explosions following the
gravitational collapse of the cores of massive (ZAMS mass $\geq
8$ \M) stars that retain part of their H envelopes. SNe II of the
subtype called "plateau" (SNe IIP) show constant luminosity lasting
from 30 days to 3-4 months. Other SNII, presenting a steep linear decline
over the same period, are named "linear" \citep[SNe IIL, ][]{barbon}.  SNe
IIP were the subject of extensive analysis by \citet{hamuy} who
pointed out a continuum in their properties and revealed several
relations between physical parameters. An independent analysis,
extending the sample to both low and high luminosities, was performed by
\citet{pphdt} who confirmed the above results.  These studies show
that most sit in the region of M$_{ej} \approx 12 - 17$ \M, regardless
of the mass of radioactive \ni, and only for relatively large \ni\/
masses ($\geq10^{-2}$ \M) M$_{ej}$ increases with M$_{Ni}$. This
suggests a possible bi-modal distribution of M$_{ej}$ versus other
observables.  Thanks to the direct identification of a few SN
precursors in deep pre-explosion images, recent studies indicate that
most SNe IIP originate from the explosion of stars with M $\lesssim$
15 M$_{\odot}$\citep{smart}.

Although SNe IIP are possibly the best studied core collapse
explosions, there are several issues that remain unclear, e.g. the
nucleosynthesis yields, the nature and rate of the mass loss in the
latest stages of evolution of the precursor (this can affect the
optical and radio display once the wind material is overcome by the
ejecta, see \citet[][]{moriya} and \citet{smith}), the influence of the metallicity
on the progenitor evolution \citep[see][]{kasen} and the location and
amount of dust formation and its impact on the observables
\citep[see][]{04et}.

\od\/ is an ideal test case to address such issues.  It was discovered
on 2007 November 2.85 UT in the nearby galaxy UGC 12846 \citep{c1}.
\citet{c2} classified it as a normal SN IIP about two weeks after
explosion, and reported some similarity with the spectrum of type II
SN 1999em 10 days after explosion. On November 6, the SN was detected
by Swift with UVOT, although not with XRT  \citep[3$\sigma$ upper limit $<$ 1.4$\times$10$^{-13}$ erg cm$^{-2}$ s$^{-1}$,][]{a1}.

\begin{figure*}
\includegraphics[width=18cm]{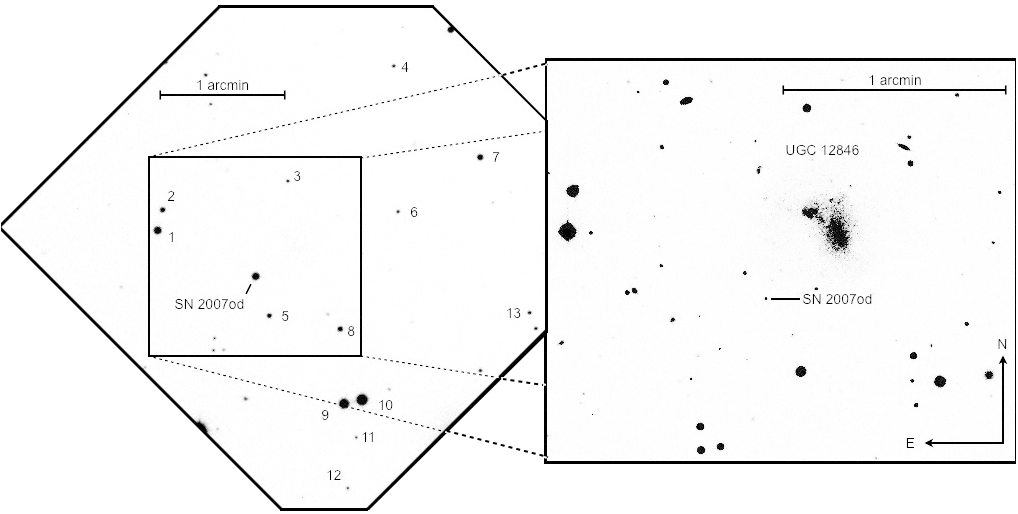}
\caption{(Left) R band image of SN 2007od in UGC 12846 obtained in November, 2007 with the 1.82-m Copernico telescope + AFOSC at Mt. Ekar (Asiago, Italy). The exposure is so short that the low surface brightness parent galaxy is barely visible. The sequence of stars in the field used to calibrate the optical and NIR magnitude of SN 2007od is also labelled. 
  (Right) Blow-up of the region of the parent galaxy in an R-band
  frame obtained on July 16, 2008 with the 2.2-m telescope of Calar
  Alto equipped with CAFOS (see Tab.~\ref{table:cht}).}
\label{fig:sn07od}
\end{figure*}

The coordinates of SN 2007od have been measured on our astrometrically
calibrated images at two different epochs: $\alpha$ =
23$^{h}$55$^{m}$48$^{s}$.68 $\pm$ 0.1$^{s}$ and $\delta$ =
+18$^{o}$24'54".8 $\pm$ 0.1" (J2000). The SN is located in a
peripheral region of UGC 12846, 38" East and 31" South of the core of
the galaxy (Fig.~\ref{fig:sn07od}). This position, slightly revised
with respect to previous determinations \citep{c1}, corresponds to a
linear distance of $\sim$6 kpc from the nucleus (assuming a distance
to UGC 12846 of $\sim$26 Mpc; cfr. Sec.~\ref{sec:red}). A prior work
based on late-time observations of \od\/ has been done by \citet{07od}.

In this paper we present optical and near-infrared observations of
\od\/ spanning from November 2007 to September 2008.  In
Sect.~\ref{sec:od} we describe photometric observations and
reduction, we estimate the interstellar reddening toward the SN site, and
describe the photometric evolution.  In Sect.~\ref{sec:spec} we
analyze the spectroscopic data. In Sect~\ref{sec:df} we provide
evidence for dust formation in SN 2007od. Discussion and conclusions
follow in Sect.~\ref{sec:dis} and~\ref{sec:final}, respectively.

\section{Photometry}\label{sec:od}
The optical and NIR photometric follow up of SN 2007od began on
November 6, 2007, four days after the discovery, and lasted until late
February 2008. Observations were resumed in June 2008 after the seasonal
gap.

\subsection{Optical ground-based data}\label{sec:opt}
Optical (UBVRI) photometry was obtained using a number of ground-based
telescopes (Tab.~\ref{table:cht}). Note that the $i$-band filter used
at the 2.56-m Nordic Optical Telescope (NOT) on January 13, 2008 is an
interference filter with central wavelength (7970\AA\/), slightly
different from the classical Gunn or Cousins I and more similar to
Sloan $i$. In our analysis, however, it was calibrated as Cousins
I.

Optical data were reduced following standard prescriptions in
IRAF\footnote{Image Reduction and Analysis Facility, distributed by
  the National Optical Astronomy Observatories, which are operated by
  the Association of Universities for Research in Astronomy, Inc,
  under contract to the National Science Foundation.} environment.
Instrumental magnitudes were measured on the images obtained
after removal of the detector signature (overscan, bias and flat field
corrections, and trimming).

Photometric zero-points and colour terms were computed for all nights
through observations of Landolt standard fields \citep{landolt}.
Eleven out of twenty-eight nights turned out to be photometric. Using
these nights we calibrated the average magnitudes of the local
sequence stars shown in Fig.~\ref{fig:sn07od}.  Magnitudes are
reported in Tab.~\ref{table:ls} along with their r.m.s. (in brackets).
For stars in common with \citet{07od}, the magnitudes differ on
average by $\Delta$B $\sim0.07\pm0.02$, $\Delta$V $\sim0.03\pm0.02$,
$\Delta$R $\sim0.05\pm0.01$, $\Delta$I $\sim0.03\pm0.02$. These
differences are probably related to the uncertainties in the
transformation to the Johnson-Cousins system adopted in \citet{07od}.
Eventually, 
the local sequence stars were used to calibrate the photometric
zero-points obtained in the non-photometric nights.

Calibrated optical magnitudes of the SN are reported in
Tab.~\ref{table:snm} along with early magnitudes from \citet{c1}.
There is no evidence of the SN in images obtained on September 17,
2008 and the values reported are the limiting magnitudes computed
by placing artificial stars near the position of the
SN. 

\begin{figure*}
\includegraphics[width=18cm]{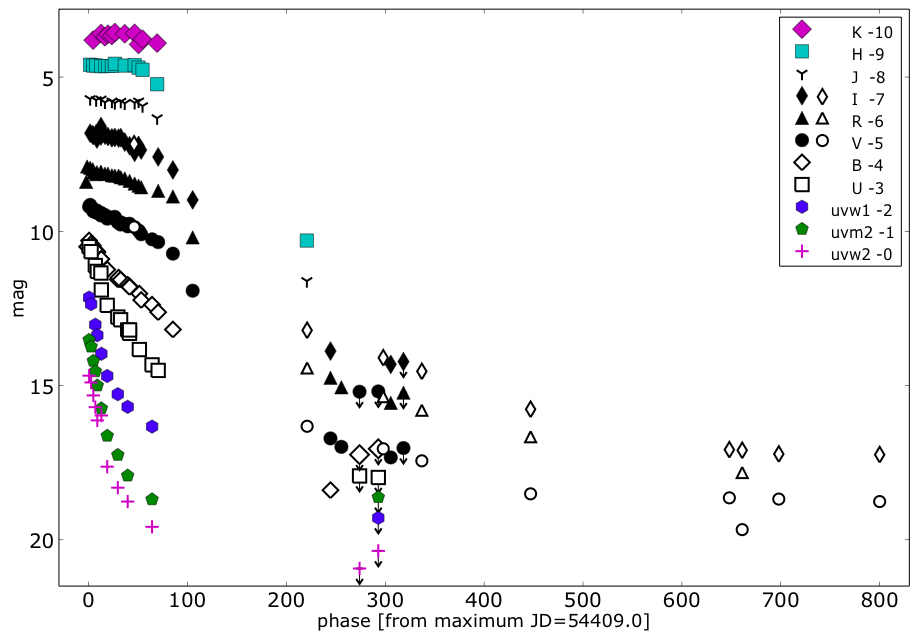}
\caption{Synoptic view of the light curves of SN 2007od  in all available bands. U, B and V light curves include data both from ground-based telescopes (Tab.~\ref{table:cht}) and SWIFT (see Sect.~\ref{sect:swift}), RIJHK light curves from ground-based telescope images, and uvw2, uvm2 and uvw1 light curves only from SWIFT. The magnitude shifts from the original value reported on Tab.~\ref{table:snm} are in the legend. Open VRI symbols  are magnitude values reported in \citet{07od}.} 
\label{fig:sn_lc}
\end{figure*}

\begin{table*}
\caption{Instrumental configurations used for the photometric follow up.}
\begin{center}
\begin{tabular}{lclclccl}
\hline
\hline
Telescope & Primary mirror & Camera & Array & CCD & pixel scale & field of view & filters \\
 & m & & &  & arcsec/pix & arcmin & \\
 \hline
 Copernico & 1.82 & AFOSC & 1024 x 1024 & TK1024AB & 0.46 & 8.1 & Bessell B V R, Gunn i\\
 TNG & 3.58 & DOLORES$^{*}$ & 2048 x 2048 & EEV 42-40 & 0.25 & 8.6 & Johnson U B V, Cousins R I\\
 				&  & NICS & 1024 x 1024 & HgCdTe Hawaii & 0.25 & 4.2  & J H K\\
REM & 0.60 & ROSS & 1024 x 1024 & Apogee Alta & 0.575 & 10  & Johnson V, Cousins R I\\
 				&  & REMIR & 512 x 512 & Hawaii I & 1.2 & 10  & J H K\\
 LT & 2.0 & RATcam & 2048 x 2048 &  EEV 42-40 & 0.13 & 4.6 & Sloan u, Bessell B V \\
 NOT & 2.5 & ALFOSC & 2048 x 2048 & EEV 42-40 & 0.19 & 6.4 & Johnson U B V R, interference i \\
 CAHA & 2.2 & CAFOS & 2048 x 2048 & SITe & 0.53 & 16 & Johnson B V R I\\
SWIFT& 0.3 & UVOT & 2048 x 2048 &  microchannel & 0.48 & 17 & uvw2,uvm2,uvw1,u,b,v \\
 & & & &intensified CCD & & &\\
 \hline
\end{tabular}
Copernico = Copernico Telescope (Mt. Ekar, Asiago, Italy); TNG = Telescopio Nazionale Galileo (La Palma, Spain); REM = Rapid Eye Mount Telescope (La Silla, Chile); LT = Liverpool Telescope (La Palma, Spain); NOT = Nordic Optical Telescope (La Palma, Spain); CAHA = Calar Alto Observatory (Sierra de los Filabres, Andalucia, Spain); SWIFT by NASA.
\end{center}
$^{*}$ Observations from 1$^{st}$ November to 31$^{st}$ December 2007 were performed with an engineering CCD with the same general characteristics.
\label{table:cht}
\end{table*}%

SN magnitudes were measured by means of the point spread function (PSF)
fitting technique.  The adoption of the PSF technique is justified by
the occurrence of the SN in the outskirts of the parent galaxy, in a
region with a smooth background. The uncertainties reported in
Tab.~\ref{table:snm} were estimated by combining in quadrature the
errors of the photometric calibration and the errors of the PSF
fitting through artificial stars.

Optical observations with the 0.6-m Rapid Eye Mount (REM) telescope plus
ROSS were affected by random deformations of the PSF. For this reason
the SN magnitudes were always determined with aperture photometry with
respect to a few local sequence stars close to SN.  REM observations
at airmasses $>2$ were rejected, because the star shapes on the frames
were elliptical and neither this approach nor PSF fitting produced
consistent measurements.

\begin{table*}
  \caption{Magnitudes of the local sequence stars in the field of SN 2007od (cfr. Fig.~\ref{fig:sn07od}). The r.m.s. of the measurements are in brackets.}
 \begin{center}
  \begin{tabular}{cccccc}
  \hline
  \hline
   ID & U & B & V & R & I \\
 \hline
 1 & 15.254 (.025) & 15.206 (.026) & 14.436 (.020) & 13.994 (.019) & 13.697 (.017)  \\
 2 & 17.933 (.008) & 17.035 (.011) & 15.947 (.021) & 15.335 (.017) & 14.859 (.021) \\
 3 & 18.831 (.018) & 18.238 (.010) & 17.264 (.026) & 16.798 (.033) & 16.291 (.023) \\
 4 &  & 17.833 ( - ) & 17.002 (.034) & 16.542 (.044) & 15.942 (.014) \\
 5 & 18.576 (.041) & 17.540 (.022) & 16.426 (.018) & 15.764 (.021) & 15.228 (.022) \\
 6 & 18.338 (.046) & 18.264 (.029) & 17.102 (.025) & 16.803 (.021) & 16.295 (.027) \\
 7  & 15.612 ( - ) & 15.770 (.055) & 15.106 (.023) & 14.819 (.029) & 14.373 (.020)  \\
 8 & 18.351 (.009) & 17.324 (.018) & 16.122 (.024) & 15.438 (.020) & 14.760 (.021)\\
 9 & 15.501 (.020) & 14.583  (.003) & 13.452 (.020) & 12.834 (.017) & 12.192 (.016) \\
 10 & 14.584 (.027) & 13.845 (.006) & 12.858 (.017) & 12.304 (.019) & 11.743 (.018) \\
 11 & 18.522 (.011) & 18.271 (.031)& 17.463 (.026) & 17.090 (.029) & 16.608 (.014)\\
 12 & 17.748 ( - ) & 17.955 (.030) & 17.309 (.042) & 17.058 (.031)& 16.644 (.022)\\
 13 &   & 17.926 ( - ) & 16.777 (.040) & 16.079 (.013) & 16.005 (.002) \\
 \hline
\end{tabular}
\end{center}
\label{table:ls}
\end{table*}

\begin{table*}
\caption{Ground-based UBVRI magnitudes of SN 2007od and assigned errors (in brackets).}
\begin{center}
\begin{tabular}{cccccccc}
\hline
\hline
Date & JD & U & B & V & R & I & Inst.\\
dd/mm/yy & (+2400000) & & & & & &\\
\hline
02/11/07 & 54407.35 &       &      &      & 14.4   ( - )  &      &   7  \\
03/11/07 & 54408.30   &   &  14.5  ( - )   &    &  13.9   ( - )  &    &     7   \\
06/11/07 & 54411.57  &    &    &  14.144 (.025)  &  13.945 (.022)  &  13.821 (.024)	&  2\\
08/11/07 & 54413.21 &    &   14.464 (.028) & 14.263 (.026)   & 13.991 (.021)  &  13.873 (.023)	&  1\\
09/11/07 & 54413.62 &   &     &  14.350 (.027)  &  14.095 (.025)  &  13.893 (.028)	 & 2\\
12/11/07 & 54416.64  &    &    &  14.392 (.032)  &  14.148 (.028)  &  13.846 (.024)	 & 2\\
12/11/07 & 54417.42  &     &   14.655 (.029)  & 14.329 (.025)  &   14.084 (.021)  &  14.009 (.022)   &     1	\\
16/11/07 & 54421.43  & 14.353 (.033)   &   14.902 (.027) & 14.486 (.031)  &  14.069 (.028)  &  13.913 (.023) & 3    \\
16/11/07 & 54421.54  &   &    & 14.415 (.027)  &  14.107 (.025)  &  13.606 (.024)	&  2\\
21/11/07 & 54425.53  &     &     & 14.474 (.050)  &  14.132 (.042)  &  13.886 (.031)	&  2  \\
24/11/07 & 54428.58 &    &    &14.535 (.046)  &  14.135 (.050)  &  13.930 (.055)	&  2\\
27/11/07 & 54432.54  &     &    & 14.558 (.028)   & 14.186 (.027)  &  13.971 (.025)	&  2\\
01/12/07 & 54435.56  &     &    &  14.533 (.028)    & 14.194 (.026)   & 13.960 (.025)	 & 2\\
05/12/07 & 54439.34  &     &   15.503 (.034)  &14.707 (.025)  &  14.198 (.023)  &  13.980 (.022) &	  1	\\
06/12/07 & 54441.35  & 15.868 (.036)   &   15.570 (.030) & 14.780 (.025)  &     &        &   4	\\
07/12/07 & 54441.55  &    &   &  14.709 (.034)  &   14.249 (.029)  &  13.967 (.025)	&  2\\
11/12/07 & 54445.55  &     &     & 14.764 (.034)  &   14.288 (.028)  &   14.108 (.025)	&  2\\
15/12/07 & 54450.34  & 16.308 (.038)   &   15.784 (.028) & 14.803 (.025)  &     &     &       4\\	
15/12/07 & 54450.45  & 16.198 (.038)  &    15.805 (.026) & 14.756 (.025)  &  14.367 (.021)   & 14.228 (.025) & 3	\\
21/12/07 & 54455.55  &    &      & 14.870 (.085)  &  14.452 (.021)  &  14.440 (.029)	&  2\\
25/12/07 & 54459.53  &    &    & 14.975 (.127)  &  14.522 (.028)  &  14.217 (.029)	&  2\\
25/12/07 &  54460.33 &  16.833 (.038)   &   16.022 (.028) & 15.006 (.028)  &    &       &     4	\\
28/12/07 & 54462.20  &    &   16.228 (.028) & 15.096 (.027)  &  14.574 (.023)  &  14.366 (.022)   &      1	\\
29/12/07 & 54463.53 &       &    & 15.023 (.062)  &  14.454 (.028)  &  14.234 (.024)	&  2	\\
13/01/08 & 54479.36  & 17.509 (.044)    &  16.622 (.027) & 15.350 (.025) & 14.688 (.022)  &  14.589 (.021)   &  	  6	\\
28/01/08 & 54494.26 &     &   17.184 (.029) & 15.724 (.028)  &  14.871 (.023)  &  15.011 (.023)   &      1  \\
17/02/08 & 54514.25 &    &    & 16.926 (.032)  &  16.197 (.025)  &  15.983 (.028) &	  1\\	
05/07/08 & 54653.62  &    &   22.392 (.197) & 21.714 (.164)  &  20.748 (.058)  &  20.882 (.264) &	  5	\\
16/07/08 & 54664.62 &    &    &  21.987 (.101)  &  21.057 (.064)  &   	&  5	\\
04/09/08 & 54714.59  &  &   & 22.330 (.057)  &  21.564 (.028)  &  21.314 (.032)	&  3	\\
17/09/08 & 54727.42 &     &    & $>$ 22.0  & $>$ 21.2   & $>$ 21.2 	&  5	\\
\hline
\end{tabular}
\end{center}
1 = Copernico,  2 = REM, 3 = TNG, 4 = LT, 5 = CAHA, 6 = NOT, 7 = CBET 1116. Telescope abbreviations are the same as in Tab.~\ref{table:cht}. 
Unfiltered photometry from CBET 1116 is considered most similar to the R-band Johnson-Bessell system.
\label{table:snm}
\end{table*}%

\subsection{Near Infrared data}
Near-infrared (NIR) photometry (JHK) was obtained with NICS mounted at
the 3.5-m Telescopio Nazionale Galileo (TNG) and with REMIR at REM
(cfr. Tab.~\ref{table:cht}).  The reduction of NIR images included the
subtraction of sky images obtained by combining several 
dithered exposures.  Then NIR images of the SN field for each band
were obtained by combining several sky-subtracted, dithered
exposures.  Photometric calibration was achieved relative to 2MASS
photometry of the same local sequence stars used for the calibration
of the optical photometry.  The NIR magnitudes of SN 2007od are listed
in Tab.~\ref{table:snir}. The values reported for the K band of NICS
were obtained with the K' filter, instead.

\begin{table*}
\caption{JHK magnitudes of SN 2007od and assigned errors (in brackets). We accounted for both measurement errors and uncertainties in the photometric calibration.}
\begin{center}
\begin{tabular}{cccccc}
\hline
\hline
Date & JD & J & H & K & Inst.\\
dd/mm/yy & (+2400000) & & & &\\
\hline
 06/11/07 &  54410.57	&   13.712 (.045)  &  13.602 (.070)	&        &   2	\\
 09/11/07 &  54413.62   &   	 &  13.628 (.044)	&  13.804 (.089)	&     2	\\
 12/11/07 &  54416.64	&   13.748 (.036)	&   13.625 (.048)	&   	 &    2\\	
 16/11/07  & 54421.54   &    13.725 (.046)	 &  13.649 (.041)	&  13.592 (.122)	&     2	\\	   
 21/11/07  & 54425.53	&   13.820 (.024)	&   13.641 (.023)	&  13.697 (.083)	&     2\\	
 24/11/07 &  54428.58 	&   	&   13.642 (.041)	&  13.612 (.062)	&     2\\	
 27/11/07 &  54432.54	&   13.792 (.019)	&   13.641 (.025)	&  13.653 (.048)	&     2\\	
 01/12/07 &  54435.56 	&   13.841 (.018)	&   13.571 (.046)	&  13.559 (.054)	&     2\\	
 07/12/07  & 54441.55	&   13.798 (.042)	&  	&    &   2	\\
 11/12/07 &  54445.55	&   13.851 (.029)	&   13.628 (.021)	&  13.598 (.069)	&     2\\
 21/12/07 &  54455.55	&   13.836 (.035)	&   13.613 (.031)	&  13.584 (.062)	&     2\\	
 25/12/07 &  54459.53	&   13.789 (.045)	&   13.694 (.044)	&  13.942 (.130)	  &   2\\	
 29/12/07 &  54463.53	&   13.941 (.019)	&   13.765 (.039)	&  13.773 (.067)	&     2\\	
 12/01/08 &  54478.31	&   14.333 (.037)	&   14.233 (.039)	&  13.901 (.029)	&     3\\	
 11/06/08 &  54629.73	&   19.612 (.029)	 &  19.300 (.047)	&    &    3	\\
 \hline
\end{tabular}
\end{center}
2 = REM, 3 = TNG (K' filter). The abbreviations and the numbers are the same as in Tab.~\ref{table:cht}.
\label{table:snir}
\end{table*}%

\subsection{SWIFT data} \label{sect:swift} 
For \od\/ a number of
optical and ultraviolet observations, obtained by UVOT on board the
SWIFT satellite, are available. UVOT data were obtained in
  uvw2, uvm2, uvw1, u, b, v filters \citep{swift} with spatial
resolution of about $2\arcsec$ (FWHM).
 
Twelve epochs of observations, spread over a period of about 60 days,
are available along with two late time observations at about 300 days
when the SN was below the detection threshold. We reduced these
data using HEASARC\footnote{NASA's High Energy Astrophysics
  Science Archive Research Center} software. All images for each epoch
were first co-added, then reduced following the guidelines of
\citet{swift}.  Aperture magnitudes were transformed from the
Swift system to Johnson magnitudes through the colour transformations
of \citet{li}.  By comparing the space and the ground-based SN
  magnitudes at corresponding epochs we found an average difference
(ground--space) $\Delta U \sim 0.19\pm0.03$, $\Delta B \sim
0.04\pm0.03$ , $\Delta V \sim 0.09\pm0.03$.  These corrections have
been applied to all UVOT magnitudes and the resulting magnitude values
are reported in Tab. \ref{table:swf}.

\begin{figure}
\includegraphics[width=\columnwidth]{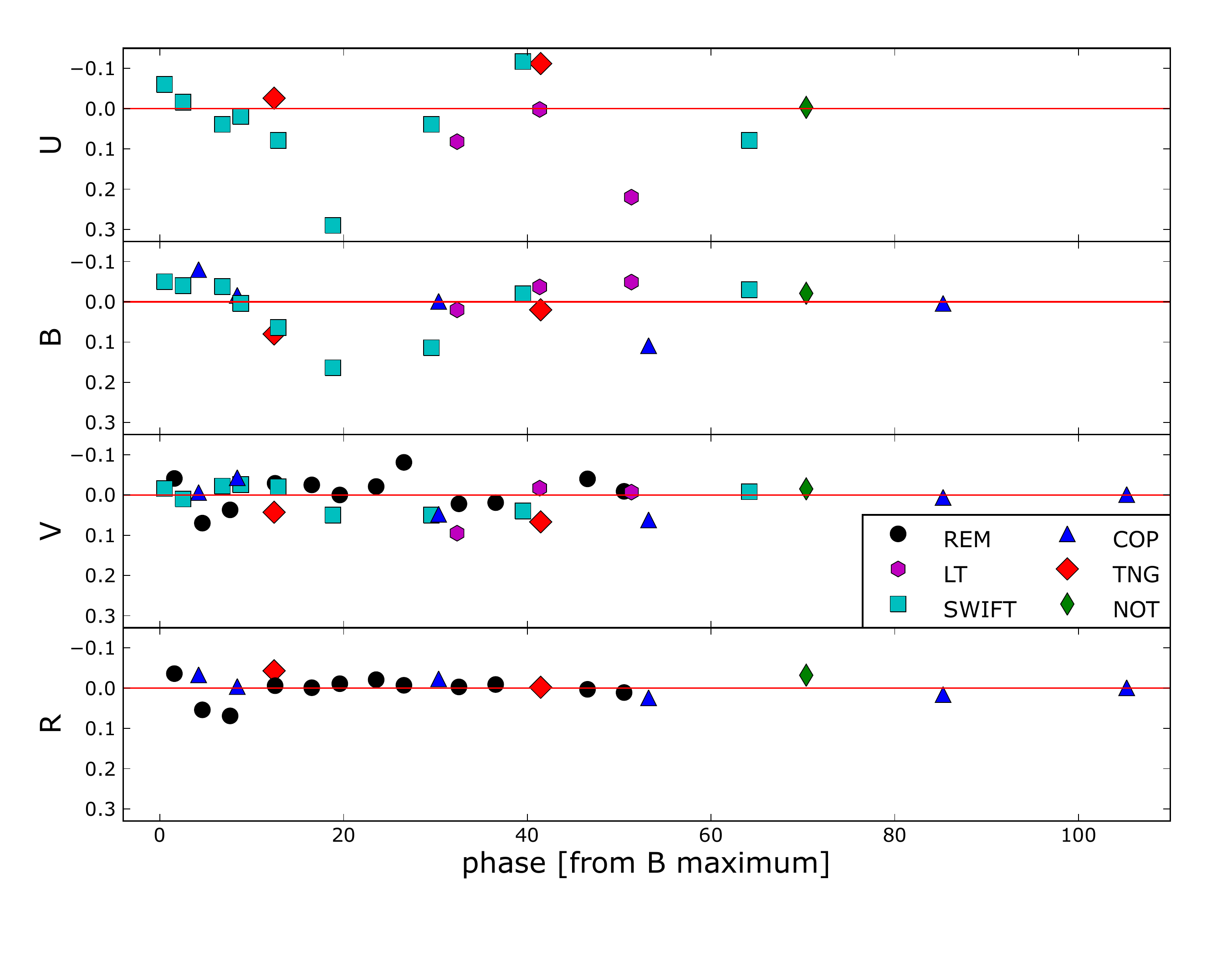}
\caption{U, B, V and R band residuals in the first 106 days with respect to low order polynomial fits of all available data. Different instruments are marked with different symbols.} 
\label{fig:res}
\end{figure}

To check for possible biases due to the peculiar pass bands of UVOT
(and other instruments), in Fig~\ref{fig:res} we show the residuals
with respect to a low order polynomial fit of all the available
values.  The agreement of the ground-based data is overall good with
few measurements deviating by more than 0.05 mag.  SWIFT data deviate
more. The largest differences are for the U band in which the specific
detector responses and the variable transmission at atmospheric
cut-off can generate quite different passbands.
       
\begin{table*}
\caption{Swift magnitudes of \od\/ and assigned errors (in brackets). The UBV magnitudes have been corrected for the small systematic differences mentioned in the text.}
\begin{center}
\begin{tabular}{cccccccc}
\hline
\hline
Date & JD & uvw2 & uvm2 & uvw1 & U & B & V\\
dd/mm/yy & (+2400000) & & & & & &\\
\hline
04/11/07 & 54409.50 &   14.68 (.05)    &   14.52 (.06)   &   14.15 (.06)   & 13.49 (.04)  &   14.34 (.06)   &   14.19 (.04)  \\
07/11/07 & 54411.52   &  14.90 (.05) &  14.74 (.06)   &  14.36 (.05)  &  13.66 (.05)  &  14.41 (.05)  &   14.22 (.04)   \\
09/11/07 & 54413.78    & 15.32 (.06) & 15.20 (.06)   &      &     &     &     \\
11/11/07 & 54415.79  & 15.70 (.05)   &  15.54 (.06)  &  15.03 (.06)  &  14.11 (.05)  &  14.60 (.05) &  14.35 (.04)\\
13/11/07 & 54417.80 & 16.13 (.06) &   16.00 (.06) & 15.37 (.06)  &  14.30 (.05)  &  14.68 (.05)	&  14.36 (.04)  \\ 
17/11/07 & 54421.88 &  15.97 (.03)  &  16.74 (.07)  & 15.97 (.07)  & 14.90 (05)  &  14.90 (.05)	&  14.41 (.04)\\
23/11/07 & 54427.83 &  17.63 (.07) &   17.63 (.08)  &  16.69 (.07)  &  15.40 (.05)  &  15.24 (.05) & 14.59 (.04)\\
04/12/07 & 54438.55 &  18.31 (.08)  &  18.25 (.11) &  17.28 (.08)  &  15.78 (.05) &  15.56 (.05)    &14.71 (.04)\\	
14/12/07 & 54448.52  &  18.76 (.11)  &  18.92 (.13) &  17.69 (.08) &  16.20 (.06) &  15.74 (.05) & 14.84 (04)\\
07/01/08 & 54473.16  & 19.58 (.13)   &   19.69 (.22) & 18.33 (.10)  &  17.33 (.07) &  16.38 (.05) & 15.25 (.04)    \\
04/08/08 & 54683.02 & $>$ 20.9 & & &$ >$ 20.9 & $>$ 21.2 & $>$ 20.2 \\
23/08/08 & 54701.96 & $>$ 20.4 & $> $19.6 & $>$ 21.3 & $>$ 21.0 & $>$ 21.0 &  $>$ 20.2 \\
\hline
\end{tabular}
\end{center}
\label{table:swf}
\end{table*}%

\subsection{Photometric evolution}\label{sec:pe}
uvw2, uvm2, uvw1, U, B, V, R, I, J, H, and K light curves of
SN 2007od are plotted in Fig.~\ref{fig:sn_lc}.  This figure includes
also photometric data from \citet{07od}.  As stressed by these
authors, the F606W and the Gemini $r$ filters (transformed respectively
to the V and R) include the \Ha\/ line, and care should be used
when comparing these data with standard photometry of other SNe.  Both
in the photospheric and in the nebular phases the data of \citet{07od} are in
good agreement with our data (cfr. Sect.~\ref{sec:opt}).

The R-band light curve shows a fast rise to the peak, estimated to
occur around JD 2454409.0$\pm$2.0.  In fact, early unfiltered
observations from \citet{c1} showed that the SN was still rising to
the R-band maximum at discovery (Nov 2.85 UT).  This is also
consistent with the phases derived for the first spectra of SN 2007od
through a cross-correlation with a library of supernova spectra
performed with the ''GELATO'' code \citep{avik}.  Therefore, hereafter
we will adopt JD 2454404$\pm$5 (October 30.5 UT) as an estimate for the
epoch of shock breakout.

\begin{figure}
\includegraphics[width=\columnwidth]{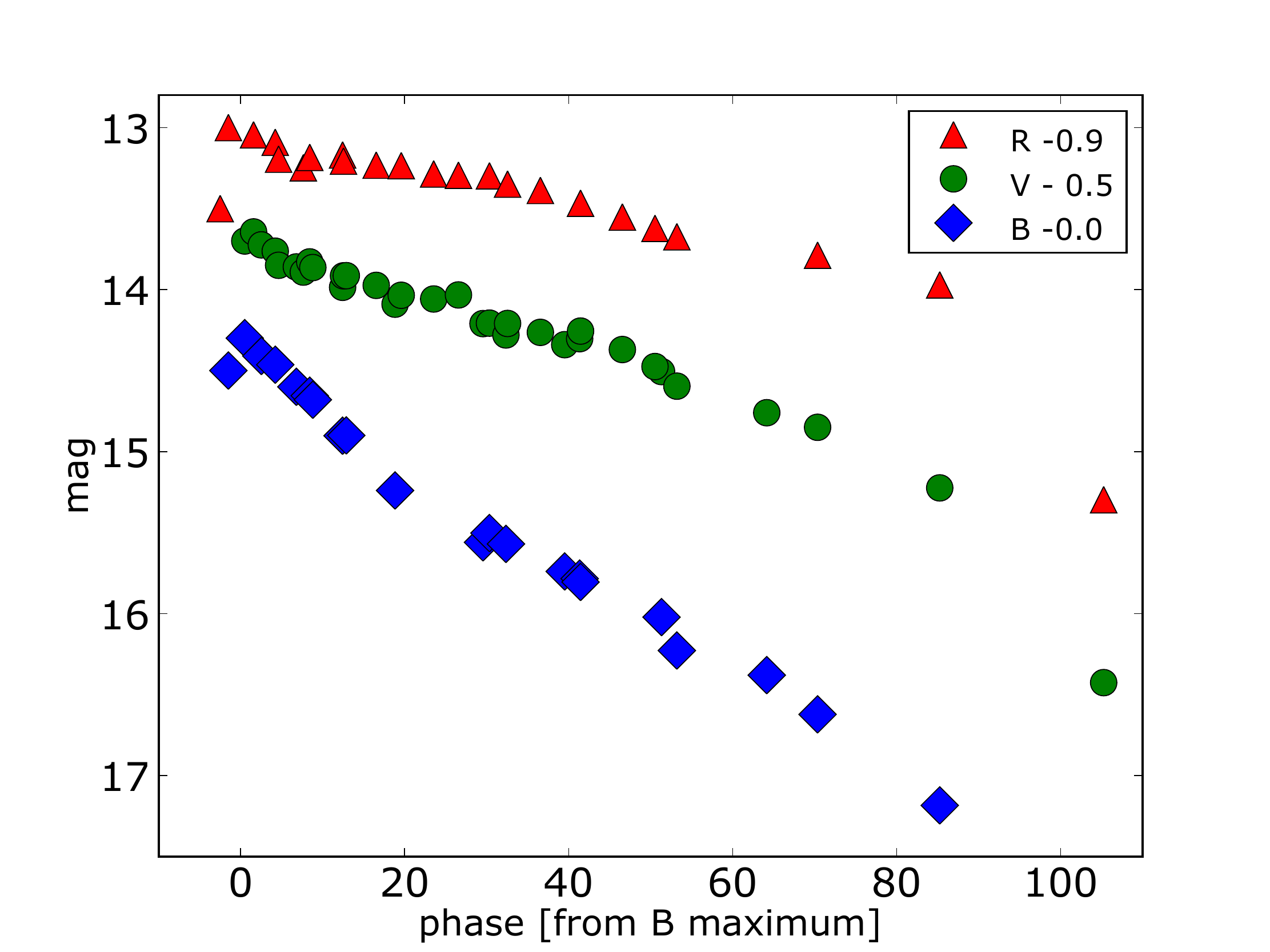}
\caption{BVR light curves in the first four months of the evolution of SN 2007od.} 
\label{fig:lcz}
\end{figure}

In Fig.~\ref{fig:sn_lc} and Fig.~\ref{fig:lcz} an early (short)
post-peak decline is visible mainly in the BVR bands, while the U band
shows a monotonic decline. 
A short, flat plateau is visible at longer wavelengths with $m_{R}
\sim 14.3$ ($M_{R} \sim -17.8$; cfr. Sec.~\ref{sec:red}) between about
day 15 and day 45. The plateau of SN 2007od is, therefore, relatively
luminous when compared with that of more typical SNe IIP
\citep{p1,richardson} and similar
to those of SNe 1992H \citep{92h} and 2004et \citep{04et,04et2}.

The SN was recovered about 240 days after maximum light.
Unfortunately, our observations do not cover the plateau-tail
transition which would allow interesting diagnostics for the explosion
parameters \citep{elm}.  It is remarkable that from the plateau to the
first point of the radioactive tail there is a drop of $\sim$6 mag in
about 200 days, which is much larger than that seen in normal SNe
IIP.  Afterwards, the late time decline rates in the different bands
(cfr. Tab.~\ref{table:mdata}), computed including also data from
\citet{07od}, are rather similar to those of most normal SNe IIP
\citep{t1,p1}. The V band decline rate is 0.94 \mcento , close to the
0.98 \mcento\/ (e-folding time 111.26 days) expected if the luminosity
is powered by the decay of \co\/ to \fe.

\begin{figure}
\includegraphics[width=\columnwidth]{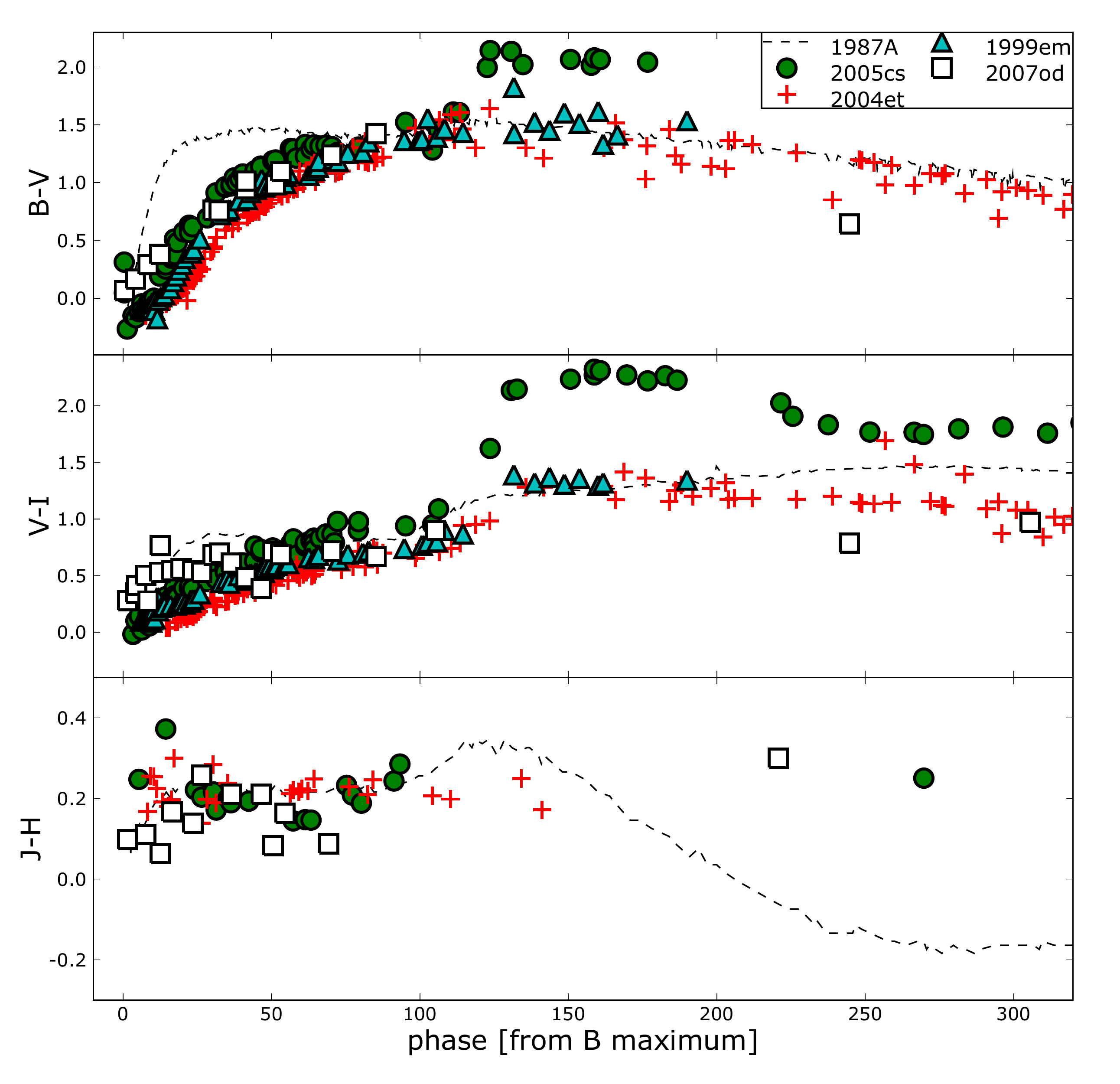}
\caption{Comparison between the dereddened colors of SN 2007od and those of  SNe 1987A, 2005cs, 1999em, 2004et. The phase of SN 1987A is with respect to the explosion date.}
\label{fig:col}
\end{figure}

In Fig.~\ref{fig:col} we show the time evolution of B--V, V--I and
J--H color curves of \od, together with those of SNe 1987A, 2005cs,
2004et, 1999em, dereddened according to the values of
Tab.~\ref{table:snc}.  All these SNe IIP show quite similar colour
evolutions with a rapid increase of the B--V colour as the supernova
envelope expands and cools down. After about 40 days the colour varies more
slowly as the rate of cooling decreases.  \od\/ follows this behavior.
The only exception to this smooth trend is SN 2005cs which shows a red
spike at about 120d (when data of SN 2007od are missing). Such a red
spike seems to mark low-luminosity SNe IIP in correspondence to the
steep post-plateau decline \citep{pa1}.  The V--I colour increases in
a linear fashion for all SNe IIP during the plateau phase, and remains
roughly constant during the nebular phase (Fig.~\ref{fig:col}). The
colour evolution of \od\/ is similar to that of other type IIP SNe,
although it is possibly redder at early phases.  In the
nebular phase, the colour curves \od\/  seem to become bluer,
especially the B--V curve, similarly to SN 2004et. 
Note, however, that such claim is based on a single point affected by large
  uncertainty.
The similarity with
2004et in the nebular phase is supported by the V--I colour evolution.
Only sparse data are available for type IIP SNe in the NIR, especially
during the nebular phase. For this reason in this domain we compared 
only a few well-studied, recent events such as SNe 2005cs and 2004et,
plus the reference type II SN 1987A. Their J--H colour curves
(Fig.~\ref{fig:col}) remain constant at J--H $\approx0.2$ mag until the phase of
$\sim$ 120 days. For the later nebular phase, the single epoch J--H
measurement of SN 2007od was found consistent with a similar
measurement for SN 2005cs.

\begin{table}
  \caption{Main parameters of type II SNe used in the comparisons with \od.}
  \begin{center}
  \begin{tabular}{ccccc}
  \hline
  \hline
  SN & $\mu$ & \ebv & Parent Galaxy & References  \\
 \hline
 1979C & 31.16 & 0.009 & NGC4321 & 1\\
 1987A & 18.49 & 0.195 & LMC & 2 \\
 1992H & 32.28  & 0.027 & NGC 5377& 3 \\
 1999em & 29.47 & 0.1 &NGC 1637& 4,5 \\
 2004et & 28.85 & 0.41 & NGC 6946 & 6 \\
 2005cs & 29.62 & 0.05 & M 51& 7 \\
  \hline
\end{tabular}
\end{center}
REFERENCES: 1 - \citet{ba80}, 2 - \citet{87a}, 3 - \citet{92h}, 
4 - \citet{99em}, 5 - \citet{99em2}, 6 - \citet{04et}, 
7 - \citet{05cs2}.
\label{table:snc}
\end{table}

\subsection{Reddening and absolute magnitude}\label{sec:red}
The Galactic reddening to UGC 12846 was estimated as
E$_{g}$(B-V)=0.038 \citep[A$_{g}(B)$=0.155, ][]{ext}.  In our best
resolution optical spectra (cfr. Sect.~\ref{sec:spec}), the absorption
features due to interstellar NaID ($\lambda\lambda$5890,5896) lines of
the Galaxy are present with average EW$_{g}$(NaID)$\sim$ 0.42
\AA\/. According to \citet{t2}, this corresponds to a galactic
reddening E$_{g}$(B-V)$\sim$0.07 mag (which can be transformed through
the standard reddening law of \citet{ca} to A$_{g}$(B)$\sim$0.28).
Interstellar NaID lines at the redshift of UGC~12846 are not detected
even in our best signal-to-noise NTT spectrum of November 27
(EW$_{i}$(NaID)$<$ 0.10 \AA) suggesting that the extinction in the
host galaxy is negligible.
This is not surprising, considering the position of the SN well
outside the main body of the galaxy, and the absence of
foreground/background structure at its location even in our late, deep
images.  Throughout this paper, therefore, for \od\/ we will assume a
total reddening of E$_{g}$(B-V)=0.038 mag.

NED provides a heliocentric radial velocity of
v$_{hel}$(UGC12846)$=1734 \pm$ 3 \kms. Correcting for the Virgo infall
\citep[$V_{Virgo}=1873 \pm 17$\kms, ][]{mould} and adopting H$_{0}$=72
\kms\/Mpc$^{-1}$, we obtain a distance modulus $\mu$=32.05$\pm$0.15
which will be used throughout this paper. This is in agreement with
the distance modulus $\mu$=31.91 (H$_{0}$=75 \kms\/Mpc$^{-1}$)
provided by \citet{tully}. 
NED provides also the recession velocities corrected for the 
Virgo Cluster, the Great Attractor, and the Shapley Supercluster
velocity fields (Virgo+GA and Virgo+GA+Shapley).
Both are marginally lower and their use would produce distance moduli smaller 
(and absolute magnitudes fainter) by about 0.1 mag.

Assuming the above distance and extinction values, we find
$M^{max}_{U} = -18.7 \pm 0.18$, $M^{max}_{B} = -17.8 \pm 0.18$,
$M^{max}_{V} = -18.0 \pm 0.23$, $M^{max}_{R} = -18.1\pm 0.21$ and
$M^{max}_{I} = -18.2 \pm 0.23$, where the reported errors include the
uncertainties of our photometry, the adopted distance modulus, and the
interstellar reddening.

\subsection{Bolometric light curve and $^{56}$Ni mass}\label{sec:bol}
\begin{figure}
\includegraphics[width=\columnwidth]{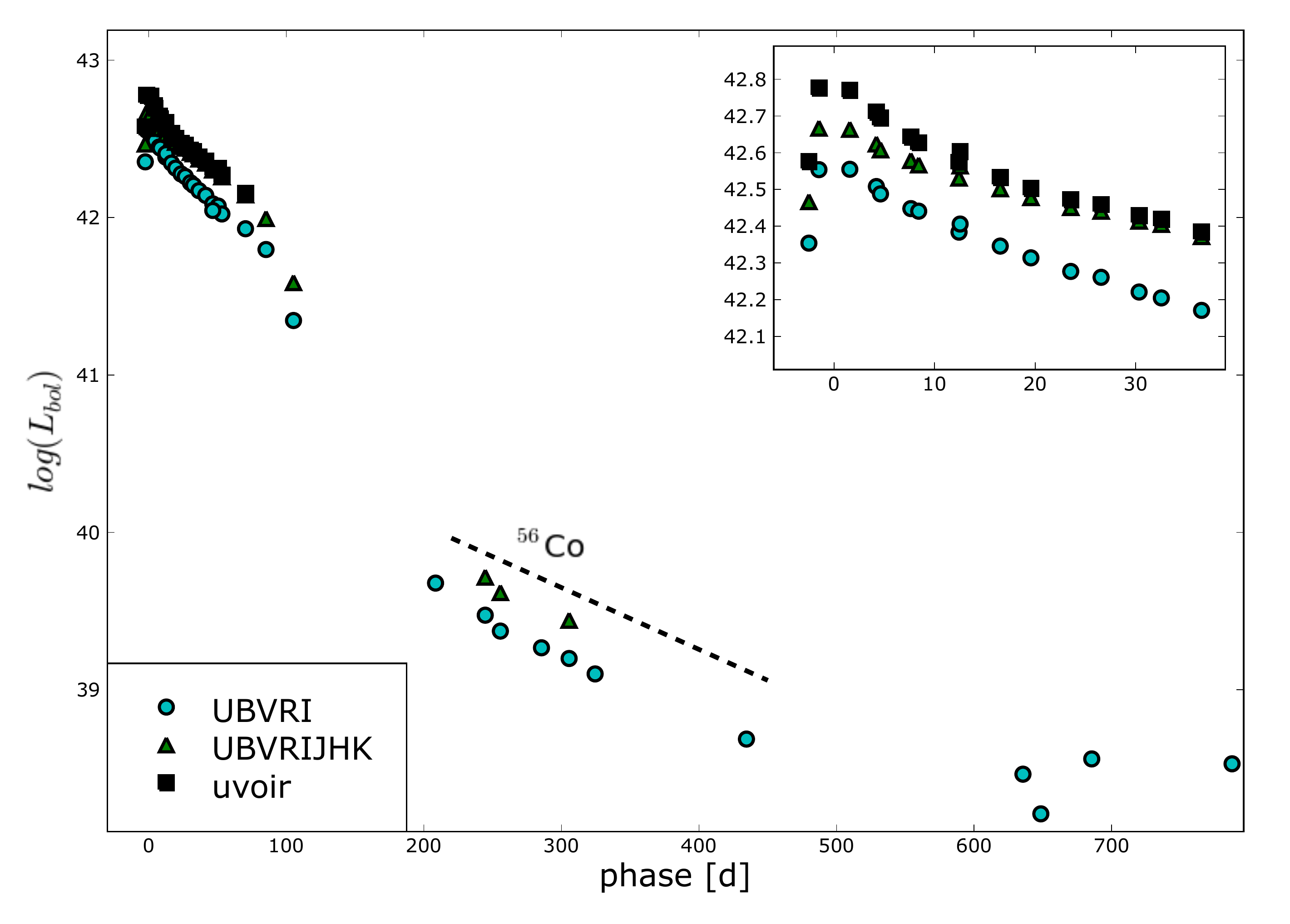}
\caption{\textit{uvoir} (black squares), UBVRIJHK (green triangles) and UBVRI (cyan circles) light curves of SN 2007od. Also reported is the extension of the UBVRI light curve to late phases obtained with data of \citet[][]{07od}. The slope of $^{56}$Co to $^{56}$Fe decay is also displayed for comparison. A blow-up until 40d post maximum is shown in the upper--right corner.  Distance modulus and reddening are those reported in Tab.~\ref{table:mdata}.} 
\label{fig:07_bol}
\end{figure}

The (uvoir) bolometric light curve of SN 2007od
(Fig.~\ref{fig:07_bol}) was obtained by converting the observed broad
band magnitudes (uvw2,uvm2,uvw1,U,B,V,R,I,J,H,K) into fluxes at the
effective wavelength, then correcting for the adopted extinction (cfr.
Sect.~\ref{sec:red}), 
and finally integrating the resulting Spectral Energy Distribution
(SED) over wavelength, assuming zero flux at the integration limits.
Flux was then converted to luminosity by using the distance adopted in
Sec.~\ref{sec:red}. The emitted flux was computed at the phases in which R
observations were available. When observations in a bandpass were
unavailable in a given night, the magnitudes were obtained by
interpolating the light curves using low-order polynomials, or were
extrapolated using constant colours.
The pre-maximum bolometric is based mainly on R band observations and
should be regarded as uncertain.  The peak of the uvoir bolometric
light curve is reached very close in time to the R maximum, on
JD$^{bol}_{max}$ = 2454410.0 $\pm$ 2.0, and at a luminosity L$_{bol}$
= 6.0 x 10$^{42}$ erg s$^{-1}$.  In Fig.~\ref{fig:07_bol}, together
with the uvoir light curve, we also plot the light curves obtained by
integrating UBVRI and UBVRIJHK contributions only.

\begin{figure}
\includegraphics[width=\columnwidth]{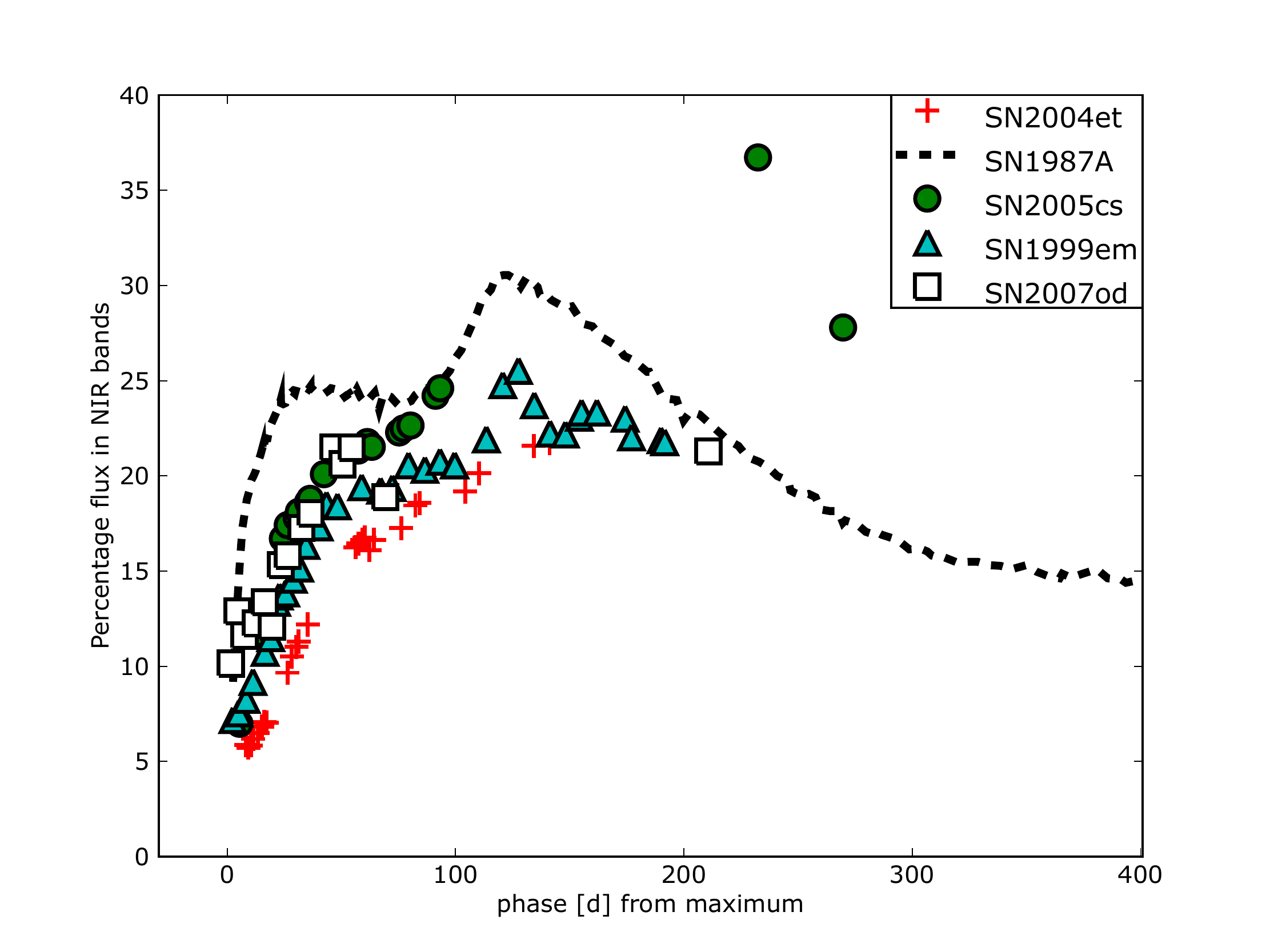}
\caption{Flux contribution of NIR bands to the U-to-K bolometric light curve for a sample of  SNe IIP.}. 
\label{fig:ir}
\end{figure}

The light curve (from U to K) of \od\/ shows a significant NIR (JHK)
contribution, as displayed in Fig.~\ref{fig:ir}.
The progressive rise of the NIR flux in the photospheric phase is
similar to that of other SNe IIP while the contribution in the nebular
phase is constant at least until $\sim$220d (Fig.~\ref{fig:ir}). This
result is similar to that found by \citet{04et} for SN 2004et.

\begin{figure*}
\includegraphics[width=18cm]{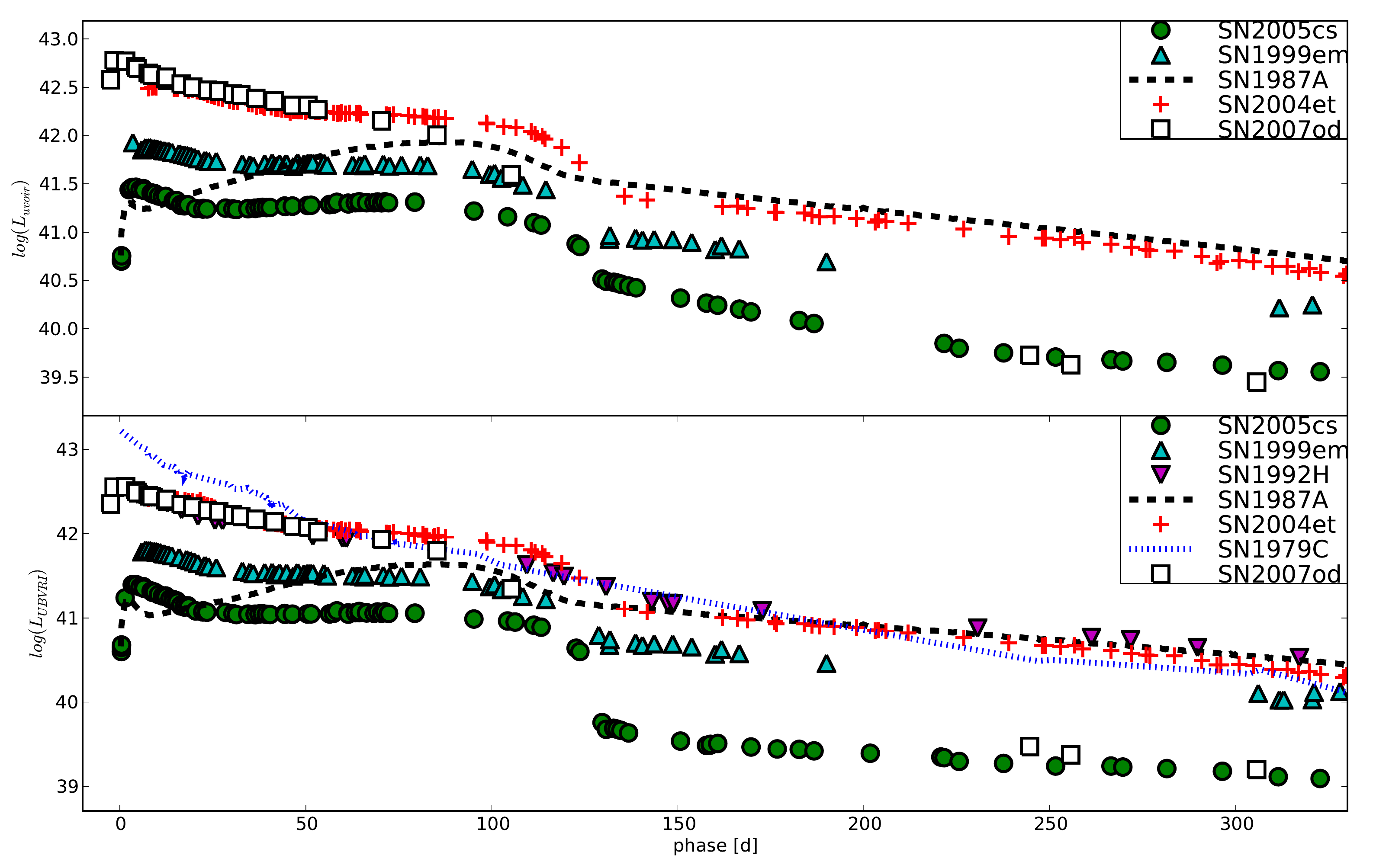}
\caption{Comparison of quasi-bolometric light curves of \od\/ with
  those of other type II SNe. The comparison on the whole
  optical-to-NIR (UBVRIJHK) domain is in the top panel, while that on
  the UBVRI range is in the bottom panel. For 
  SN 1992H the data are limited to BVR,  while for SN 1979C are UBVR.  
  The distances and reddenings adopted for the comparison with our SN
  sample are reported in Tab.~\ref{table:snc}.  Minor misalignments in
  the epoch of maxima are due to the different epochs adopted for the
  maxima of the reference band light curve and the quasi-bolometric
  curve.}
\label{fig:cfr_bol}
\end{figure*}

As for individual bands, the nebular tail of the bolometric light
curve closely matches the slope expected from the decay of $^{56}$Co
to $^{56}$Fe, suggesting complete $\gamma$-ray trapping. The measured
slope is $\gamma \sim 1.053$ mag $100d^{-1}$, close to the canonical
value of $\gamma \sim 0.98$ mag $100d^{-1}$ of \co\/ decay.  After day
500, the curve, based only on data by \citet{07od}, flattens.

In Fig.~\ref{fig:cfr_bol} we compare the bolometric light curve of
\od\/ with those of other SNe.  The comparison points out the large
drop in magnitude ($\sim$ 6 mag) from the plateau to the radioactive
tail and the small amount of $^{56}$Ni (cfr. Sec.~\ref{sec:df}). The
SNe chosen for the comparison are those reported in
Tab.~\ref{table:snc}.  Unfortunately, not many of them have spectral
coverage from the optical to the NIR, and therefore the comparison was
done either for UBVRIJHK (top panel) or UBVRI (bottom panel)
bolometric curves. The early luminosity of \od\/ is comparable to
those of the luminous SN 1992H and SN 2004et. The behavior of the light 
curve is certainly more similar to that of Type~IIP SNe with respect to the Linear SN~II~1979C 
proposed for comparison (the average decline rate by  \citet{p1} is
$\beta_{100}^{B}(07od)\sim 3.2$ \mcento, typical of SN~IIP).
The similarity to SN
1992H is noticeable in the photospheric phase where both SNe present
the same early decline and short plateau, a possible evidence of a
low-mass H envelope.  On the other hand, the radioactive tail of \od\/
is more yhan 1 dex fainter than that of SN 1992H.

The $^{56}$Ni mass ejected in the nebular phase by \od\/ can be
derived by comparing the bolometric light curve
(Fig.~\ref{fig:cfr_bol} Top) to that of \a\/ assuming a similar
$\gamma$-ray deposition fraction
\begin{equation}\label{eq:1}
M(^{56}Ni)_{07od}= M(^{56}Ni)_{87A}\times\frac{L_{07od}}{L_{87A}} M_{\odot} 
\end{equation}
where M($^{56}$Ni)$_{87A}$ = 0.075 $\pm$ 0.005 \M\/ is the mass of
$^{56}$Ni ejected by \a\/ \citep{87a1}, and L$_{87A}$ is the
bolometric luminosity at comparable epoch. The comparison gives
M($^{56}$Ni)$_{07od}$ $\sim$ 0.003 M$_{\odot}$.  Making the reasonable
assumption that $\gamma$-rays from \co\/ decay are fully
thermalized at this epoch, we crosschecked this result with the
formula
\begin{equation}\label{eq:2}
\textstyle M(^{56}Ni)_{07od} = \left(7.866\times10^{-44}\right) L_{t}exp\left[\frac{(t-t_{0})/(1+z)-6.1}{111.26}\right] M_{\odot}
\end{equation}
from \citet{hamuy}, where t$_{0}$ is the explosion epoch, 6.1d is
the half-life of $^{56}$Ni and 111.26d is the \textit{e}-folding time
of the $^{56}$Co decay, which releases 1.71 MeV and 3.57 MeV
respectively as $\gamma$-rays \citep{capdecay,w2}. This method
provides M(\ni)$\sim$ 0.003 \M, fully consistent with the previous
estimate.  Indeed Fig.~\ref{fig:cfr_bol} also shows that
L$_{07od}$ is similar to that of SN 2005cs, which ejected $\sim 0.003$
M$_{\odot}$ $^{56}$Ni \citep{05cs2}, and to other under-luminous type
II events \citep{pa1}.

The luminous plateau of \od, comparable with that of the brightest SNe
IIP, coupled to an under-luminous tail is very unusual \citep{pphdt}.
In Sect.~\ref{sec:sa} we will show that on day 310 \citep[but
also in the spectra on day 226 by ][]{07od} there is evidence of dust
formation. Thus the low luminosity in the nebular phase may be due not
only to a low \ni\/ production.  The value determined above should be
considered as a lower limit.  Indeed, the late-time MIR (mid infrared)
data published from \citet{07od} allows us to study the SED up to the
M band.  The IR bands show clear evidence of strong blackbody
emission due to the re-emission of radiation absorbed at shorter
wavelengths. 
Adding this contribution  to the bolometric flux, Eq.~\ref{eq:2} provides 
M(\ni)$_{07od} \sim 0.02$ M$_{\odot}$ at the first two epochs of MIR observations.
The estimate on a third epoch is less reliable
because based only on 3.6 and 4.5 $\mu$ observations. 
We address the issue of dust formation in Sec.~\ref{sec:dis}.

\section{Spectroscopy}\label{sec:spec}
The spectroscopic monitoring of SN 2007od was carried out with several
telescopes over a period of eleven months. The journal of
spectroscopic observations is reported in Tab.~\ref{table:sp}.

\begin{figure*}
\includegraphics[width=18cm]{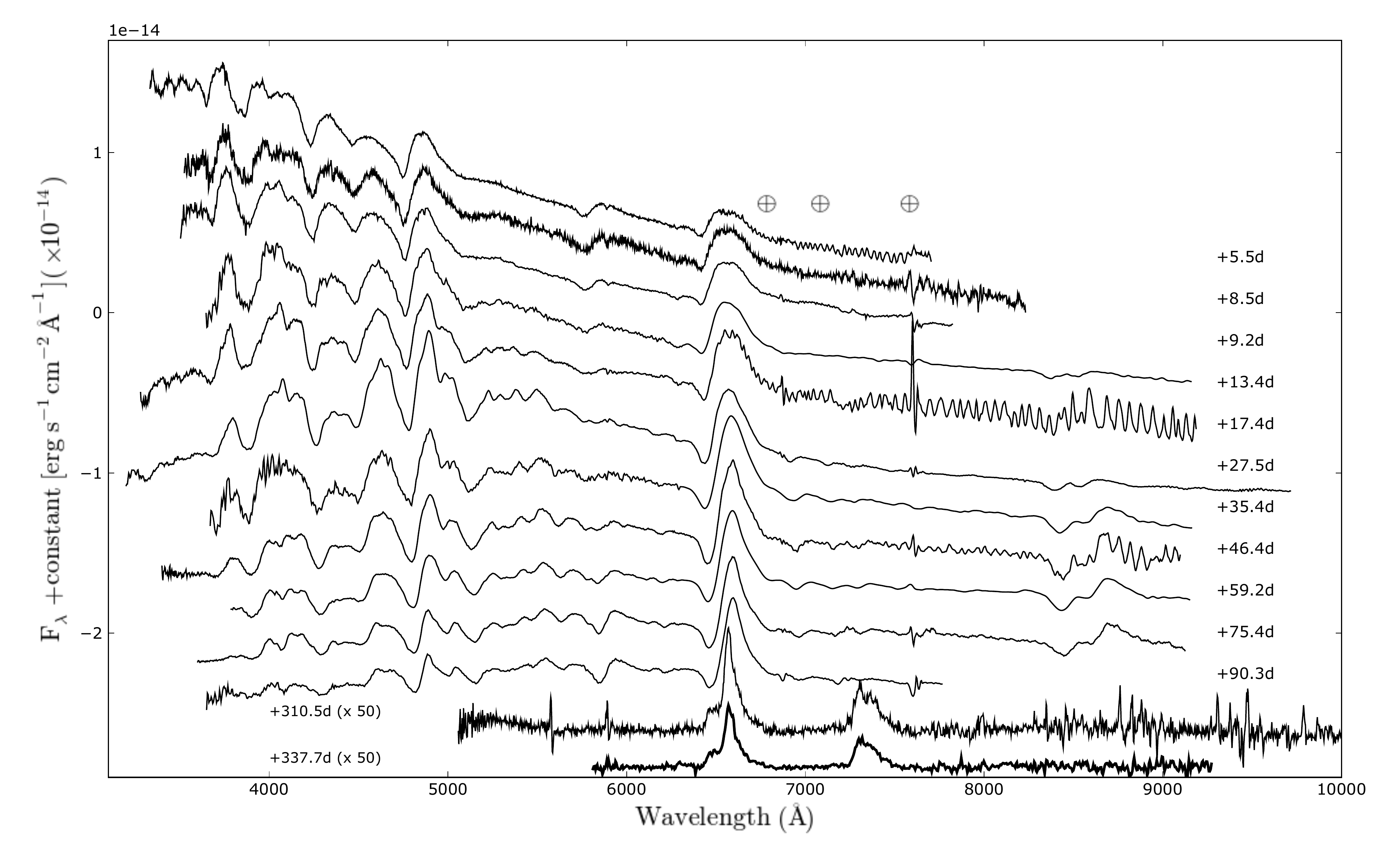}
\caption{The overall spectral evolution of SN 2007od. Wavelengths are
  in the observer's rest frame. The phase reported for each spectrum
  is relative to the explosion date (JD 2454404). The $\oplus$ symbols
  mark the positions of the strongest telluric absorptions. The
  ordinate refers to the top spectrum; the other spectra are shifted
  downwards with respect to the previous one by $2\times 10^{-15}$
  (second spectrum), $4.3\times 10^{-15}$ (third) and $2.2\times
  10^{-15}$ erg s$^{-1}$ cm$^{-2}$ \AA\/$^{-1}$(others).}
\label{fig:spec_ev}
\end{figure*}

\subsection{Spectra reduction}\label{sec:sr}
Spectra were reduced (trim, overscan, bias and flat--field corrected)
using standard routines of IRAF. Optimal extraction of the spectra was
implemented to improve the signal to noise (S/N) ratio. Wavelength
calibration was performed through spectra of comparison lamps obtained
with the same configuration as the science observations. Atmospheric
extinction correction was based on tabulated extinction coefficients
for each telescope site. Flux calibration was performed using
spectrophotometric standard stars observed in the same night with the same set-up of the
SN. Absolute flux calibration was then checked
through comparison with the photometry. This analysis was done by
integrating the spectral flux transmitted by standard BVRI filters.
When necessary, a small multiplicative factor was applied.  The
resulting calibration error is $\pm$0.1 mag.  The spectral
resolutions reported in Tab.~\ref{table:sp} were estimated from the full width
at half maxima (FWHM) of the night sky lines.  We finally used the
spectra of the standard stars to remove (or reduce) the
telluric features in the SN spectra. The regions of the strongest
atmospheric absorptions are marked in Fig.~\ref{fig:spec_ev}.
Spectra of similar quality obtained in the same night with the same
telescope were combined to increase the S/N ratio.

The spectrum of November 8 (9.2d) revealed some problems in the flux
calibration of the blue side. The spectral continuum was forced to
follow the SED derived with a fourth order polynomial fit to the
photometry obtained in the same night. This spectrum will not be used
to estimate the continuum temperature in Sec~\ref{sec:ev}.

\begin{table*}
  \caption{Journal of spectroscopic observations of \od.}
  \begin{tabular}{cccccc}
  \hline
  \hline
  Date & JD & Phase$^{*}$ & Instrumental & Range & Resolution$^{\dagger}$ \\
  & +2400000 & (days) & setup$^{\diamond}$  &(\AA) & (\AA) \\
 \hline
 04/11/07 & 54409.5 & 5.5 & TNG+DOLORES+LRB$^{\ddagger}$ & 3330-7700 & 14  \\
 07/11/07 & 54412.5 & 8.5 & Pennar+B\&C+300tr/mm & 3520-8230 & 10\\
 08/11/07 & 54413.2 & 9.2 & Copernico+AFOSC+gm4 & 3500-7820 & 24 \\
 12/11/07 &  54417.4 & 13.4 & Copernico +AFOSC+gm4,gm2 & 3640-9150 & 39 \\
 16/11/07 & 54421.4 & 17.4 & TNG+DOLORES+LRB,LRR$^{\ddagger}$ & 3280-9190 & 14 \\
 27/11/07 & 54431.5 & 27.5 & NTT+EMMI+gm2,gr5 & 3200-9710 & 11 \\
 05/12/07  & 54439.4 & 35.4 & Copernico +AFOSC+gm4,gm2 & 3660-9160 & 25  \\
 15/12/07 & 54450.4 & 46.4 & TNG+DOLORES+LRB,LRR$^{\ddagger}$ & 3400-9100 & 14\\
 28/12/07 & 54463.2 & 59.2 & Copernico +AFOSC+gm4,gm2 & 3780-9150 & 23 \\
 13/01/08 & 54479.4 & 75.4 & NOT+ALFOSC+gm4 & 3600-9120 & 13 \\
 28/01/08 & 54494.3 & 90.3& Copernico +AFOSC+gm4 & 3640-7760& 23\\
 05/09/08 & 54714.5 & 311 & TNG+DOLORES+LRR & 5050-10250& 16\\
 02/10/08 & 54741.7  & 338 & PALOMAR+DBSP+red & 5800-9270& 17\\
 \hline
\end{tabular}
 \begin{flushleft}
$^{*}$ with respect to the explosion epoch (JD 2454404)\\
 $^{\diamond}$ the abbreviations are the same as in Tab.~\ref{table:cht} with, in addition, Pennar = 1.22m Galileo Galilei telescope (Pennar, Asiago, Italy) \\
 and PALOMAR = 5.1m Hale Telescope (San Diego County, California, U.S.A.)\\
 $\dagger$as measured from the full-width at half maximum (FWHM) of the night sky lines\\
 $\ddagger$ engineering CCD
\end{flushleft}
\label{table:sp}
\end{table*}

\subsection{Spectra analysis}\label{sec:sa}
Fig.~\ref{fig:spec_ev} shows the entire evolution of \od\/ from the
first spectrum (near the R-band maximum) to about 1 year later. The
evolution is very well sampled in the photospheric phase. As for the
photometry, the transition from photospheric to nebular stages was not
observed because the SN was behind the Sun.  Line identification
relies on qualitative arguments based on the presence of lines of the
same ions at consistent velocity and by comparison with template
spectra of other SNe IIP, e.g., 2005cs \citep{05cs}, 2004et
\citep{04et} and 1999em \citep{99em2,99em,pphdt}. For most of them
detailed line identification was carried out also with spectral
modeling.

\begin{figure}
\includegraphics[width=9.cm]{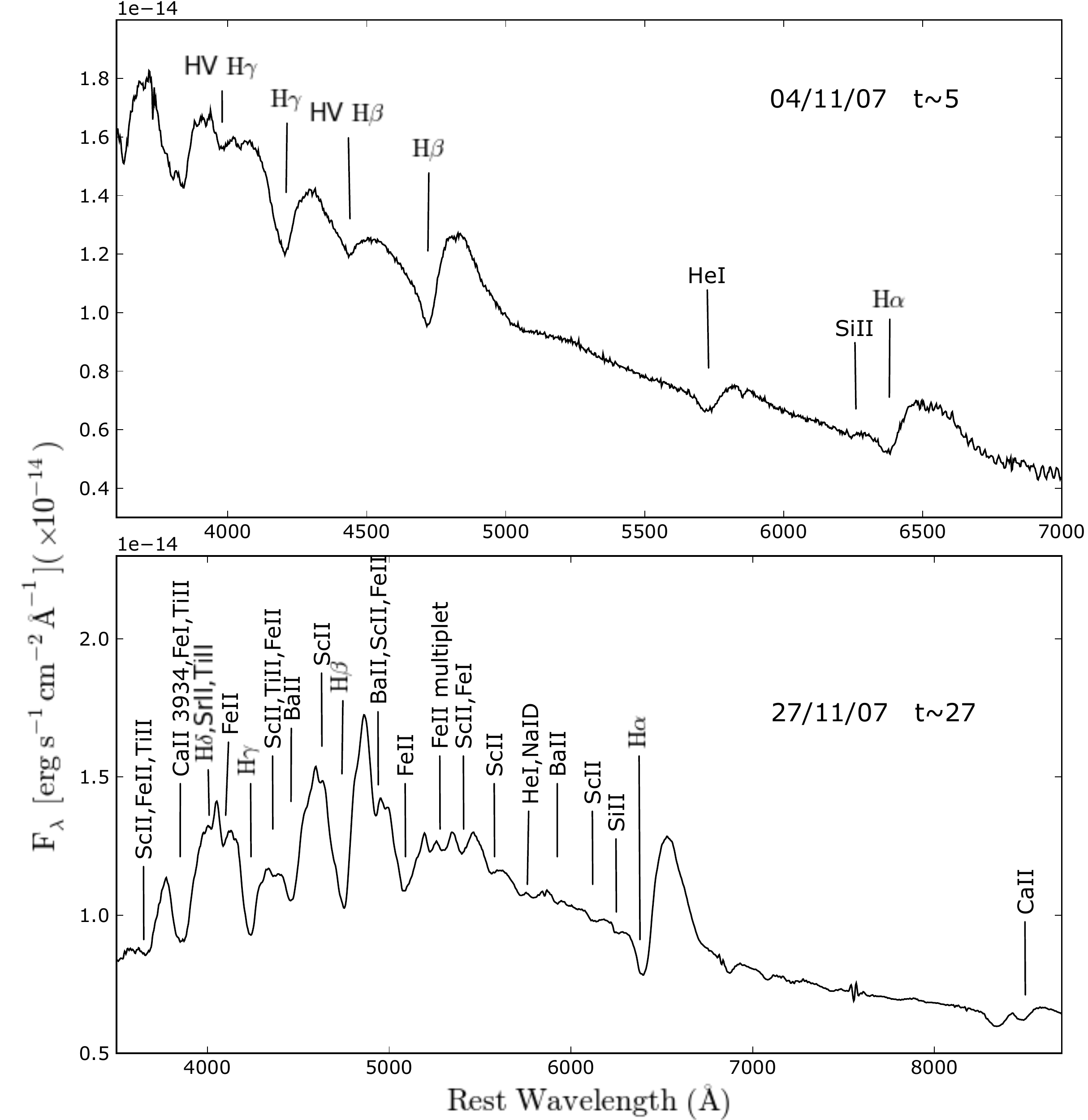}
\caption{Top: optical spectrum of SN 2007od, obtained 5 days past explosion (JD 2454404). Bottom: optical spectrum of SN 2007od, 27 days past explosion. Both spectra have been corrected for absorption in our galaxy and corrected by redshift. The most prominent absorptions are labelled.} 
\label{fig:id1}
\end{figure}

The first spectra (5-13 days) are characterized by a blue continuum,
yet cooler than those of SNe 1999em and 2005cs (cfr.
Fig.~\ref{fig:cfr_1}).  The most prominent features
(Fig.~\ref{fig:id1}, top) are the H Balmer lines and He I
$\lambda$5876, all with a normal P-Cygni profile except \Ha\/
which shows a weak absorption component and a boxy emission.  
This profile, which resembles a detached atmosphere
profile \citep{jeffery}, might be the signature of a weak interaction 
with a low density CSM
\citep[e.g. SN 1999em][]{pooley}.

Two absorption features are worth mentioning: one is prominent on the
blue side of \Hb\/ at about 4440\AA\/ (marked as high velocity,
HV, \Hb\/ in Fig.~\ref{fig:id1} Top), the other, much fainter,
on the blue side of \Ha, at about 6250\AA\/ in the first
spectrum (marked as Si II in Fig.~\ref{fig:id1} top). These lines were
noted before in the spectra of SNe IIP.  In the case of \cs\/
\citep{05cs} the feature near 4440\AA\/ was identified as N II
$\lambda$4623, supported by the presence of another N II line at about
5580\AA\/.  The latter line is not seen in the spectra of \od\/ at the
expected position.  \citet{99em2} discussed such lines in an early
spectrum of \em. While the parametrized SYNOW code suggested that the
4400\AA\/ feature was consistent with N II, the non-LTE code \texttt{PHOENIX}
rejected this identification because of the absence of the feature
around 5580\AA. Instead, the positions of both features were consistent
with being Balmer lines produced in a high--velocity (HV) layer.  Such
combined identification does not seem plausible in the case of \od\/
because the expansion velocity of the HV \Hb\/ line is much
faster than that of the putative HV \Ha\/ (25000 vs. 15000
\kms).

In our first set of spectra (5d - 13d), another absorption line is
visible around 3980\AA\/, between \Hg\/ and the H\&K doublet of
Ca II. This feature (marked as HV \Hg\/ in Fig.~\ref{fig:id1}
top) has the same velocity of the mentioned HV \Hb\/
($\sim25000$ \kms) and might therefore be produced in the same layer.
Up to phase $\sim$ 13d the trend of these lines is the same.  From day
17 the HV \Hg\/ disappears, likely hidden by the increasing
strength of metal lines such as Fe II and Ti II.  We suggest that
these are indeed HV Balmer lines and that the HV \Ha\/
component is missing because \Ha\/ level is mostly populated
collisionally, similarly to what is seen in SN IIL \citep{branch}.
This difference among the Balmer lines can be noted also in the early
spectra of SN 1993J, which showed a more pronounced P-Cygni absorption
for \Hb\/ and \Hg\/ than for \Ha.  
We also mention the possibility that a contribution to the 4440 \AA\/ and the 3980 \AA\/ features is given by HeI $\lambda$4471 and \Hd\/, respectively.
However, an extensive study with SYNOW and \texttt{PHOENIX} on the early spectra of SN 2007od \citep{norman} seems to favour the identification of the two features with HV components of \Hb\/ and \Hg\/.

In the early spectra of SN 2007od, H\&K of Ca II and a combination of
Fe I, Ti II and Sc I around 4000\AA\/  are also identified.  The line
at about 6250\AA\/ is probably Si II $\lambda$6355 as suggested by the
expansion velocity, consistent with that of the other metal ions. The
presence of Si II is fully blown at early times in type Ia, but also
proposed in several Type II SNe such as 2005cs \citep{05cs}, 1992H
\citep{92h}, 1999em \citep{99em3} and 1990E \citep{90e}.

In the spectrum of November 12 (phase = 13.4 days) a faint feature appears
at about 6200\AA\/ and is probably due to Sc II ($\lambda$6300) or Fe
II ($\lambda$6305). In this spectrum there is also evidence of a
feature at about 6375 \AA, close to the blue edge of the \Ha\/
absorption, possibly due to Fe II $\lambda$6456. The 13.4d spectrum,
the first extending to 1 micron, shows the presence of the Ca II IR
triplet ($\lambda\lambda$ 8498, 8542, 8662).

In the subsequent set of spectra (phase 17-90 days) a number of
well developed features with P-Cygni profiles appear, in addition to
the persistent Balmer lines of H (Fig.~\ref{fig:id1}, bottom).  In the
region between \Hb\/ and \Hg, a group of lines
identified as Sc II, Ti II and Fe II around 4420\AA\/, Ba II at
$\lambda$4450 and Sc II $\lambda$4670 (on the blue side of
\Hb) become more prominent than in earlier spectra. At this
phase several other metal lines appear. The Fe II multiplet 42 lines
($\lambda\lambda$ 4924, 5018, 5169) are visible to the red side of
\Hb\/ along with Fe I and Sc II in the region between 5200\AA\/
and 5500 \AA\/ and, starting from day 17, Sc II $\lambda$5658.
Other features are identified as a blend of Ba II $\lambda$4997 and Sc
II $\lambda$5031 on the red side of
\Hb\/ 
and Sc II $\lambda$6245. On day 27 the He I $\lambda$5876 line is
weak and blended with Na ID which becomes progressively stronger and
replaces He I in subsequent spectra.

The presence of C I lines in spectra of type IIP SNe was claimed
in some objects, e.g., in SNe 1995V \citep{fassia} and 1999em
\citep{pphdt}. In spectra of \od\/ at $\sim$ 75 days a faint
absorption on the red wing of Ca II IR triplet could be attributed to
C I $\lambda$9061 but no other C I lines are visible to support this
finding.

Two spectra are available in the nebular phase (310-338 days). In both
spectra the peaks of the \Ha\/ emission are blue-shifted by the
same amount ($\sim$1500 \kms). The line shows a multi--component
structure that will be discussed in Sec.~\ref{sec:df}.  These spectra
show also forbidden emission lines of [Ca II] $\lambda\lambda$7291,
7324, and weak evidence of [O I] $\lambda\lambda$6300, 6364 and [Fe
II] $\lambda$7155, which are common features in SNe
IIP.

\subsection{Comparison with other SNe}\label{sec:cfr}

\begin{figure}
\includegraphics[width=\columnwidth]{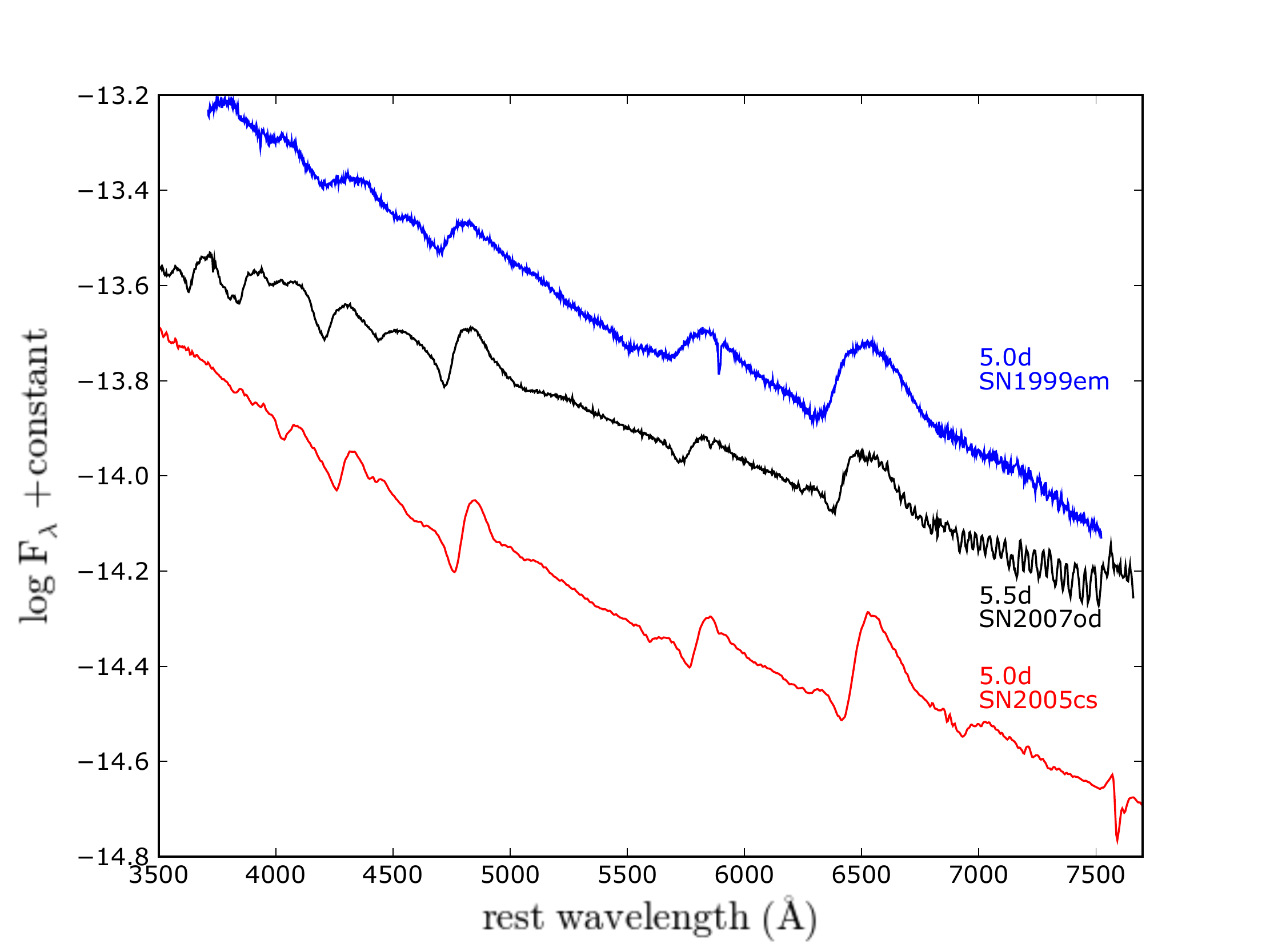}
\caption{Comparison among spectra of \em, \od \/ and SN 2005cs around 5 days after explosion. For references see Sect.~\ref{sec:cfr} and Tab.~\ref{table:snc}.} 
\label{fig:cfr_1}
\end{figure}

In Fig.~\ref{fig:cfr_1} we compare the spectrum of \od\/ at $\sim$ 5
days with two other young SNe IIP: \em\/ \citep{99em} and SN 2005cs
\citep{05cs}. 
The similarity with \em\/ was prompted by the GELATO spectral
comparison tool \citep{avik}, while the comparison with the faint SN
2005cs is made because of the characteristic features on the blue side
of \Hb\/ and Si II lines.  All spectra show relatively blue
continua and display H Balmer lines and He I 5786\AA.  The spectrum of
\od\/ 
shows a boxy profile of \Ha\/ which suggests an ejecta-CSM
interaction scenario, but no radio or X-ray observations at early phases
are available to confirm such a hypothesis.

The feature at about 4440\AA, discussed in Sec.~\ref{sec:sa}, is
possibly detected in SN~2005cs, but not in \em.  A weak line around
5600\AA\/, observed only in SN 2005cs \citep[cfr. ][]{05cs} and
(possibly) in \em, and tentatively identified as N II, is not visible
in \od.

\begin{figure}
\includegraphics[width=\columnwidth]{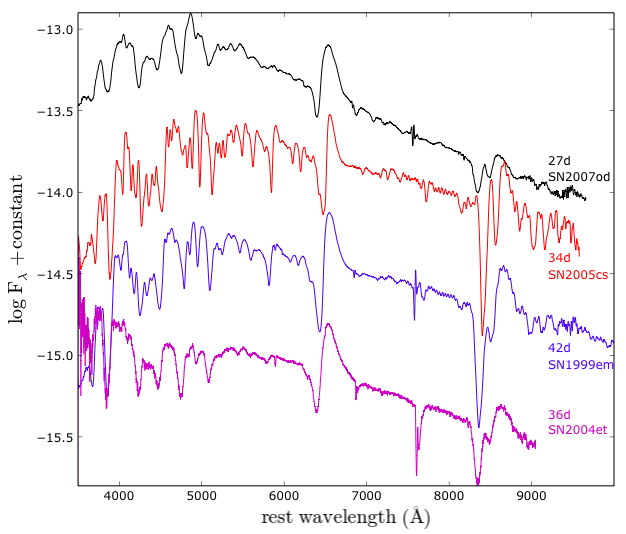}
\caption{Comparison among spectra of SN 2004et, \od, SN 2005cs and \em\/ during the plateau phase. For references see  Sect.~\ref{sec:cfr} and Tab.~\ref{table:snc}.} 
\label{fig:cfr_2}
\end{figure}

In Fig.~\ref{fig:cfr_2} a few spectra of type IIP SNe during the
plateau phase are compared. SN 2004et \citep{04et} was added here because of
the similarity in the bolometric light curve at early times
(Fig.~\ref{fig:cfr_bol}).  The spectrum of \od\/ seems to have
shallower absorption components than other SNe at the same phase,
probably because of temperature difference.  Alternatively this
effect may be due to circumstellar interaction throught
{\em\/ toplighting} effect \citep{top}. 
In such scenario the fast ejecta catches and sweeps up the much slower
circumstellar matter from the wind of the SN progenitor
(or its binary companion) and produces a continuum emission above the photosphere.
The global effect is to increase the total luminosity decreasing the contrast 
of spectral lines. It is also possible that the reverse shock decelerate the 
gas and produces low photospheric velocities.
The lines close to the blue edge of \Ha\/ at about 6300\AA\/
are visible also in the spectrum of SN 2005cs and in \em\/ but not in
SN 2004et \citep{04et}.

\begin{figure}
\includegraphics[width=\columnwidth]{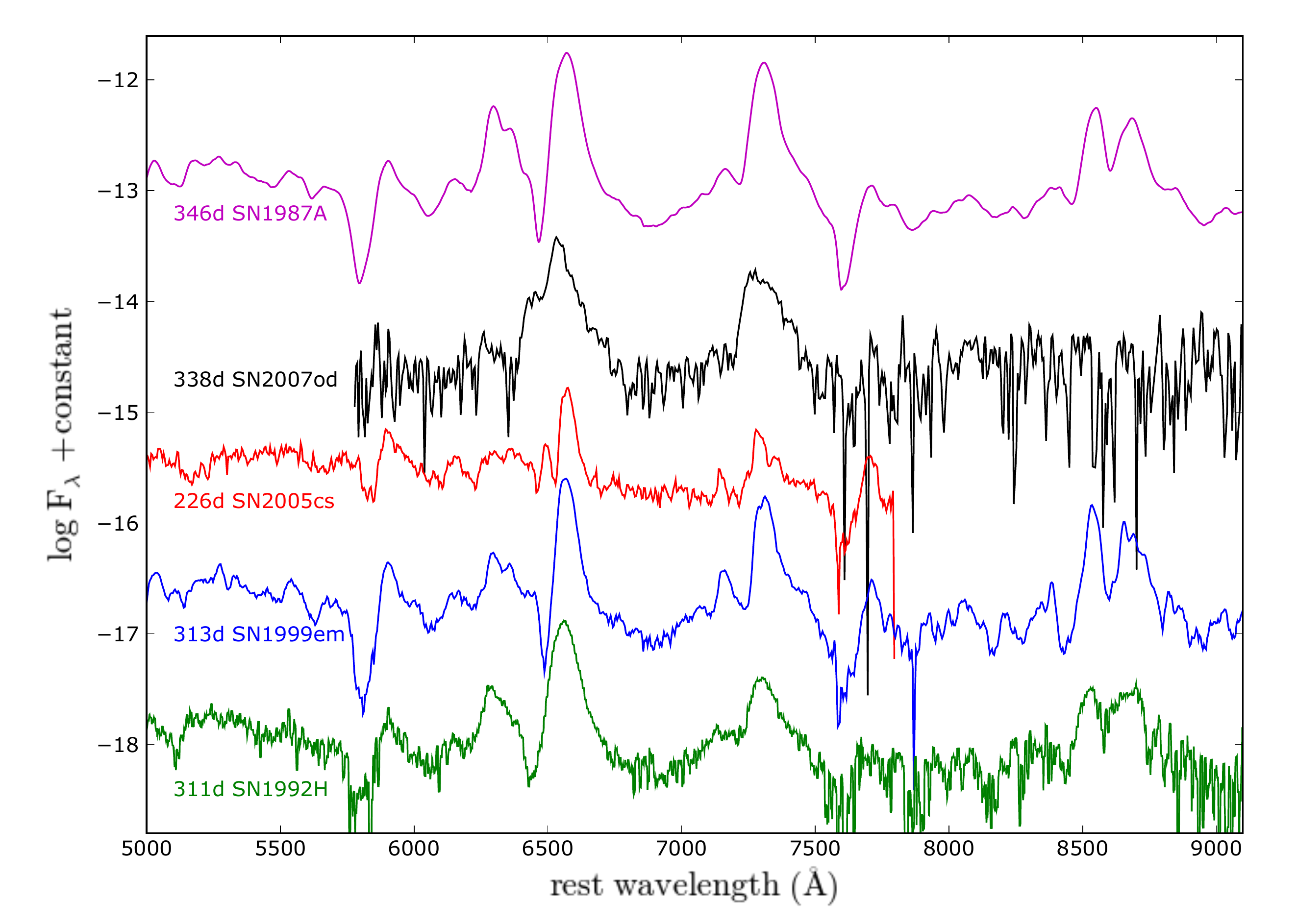}
\caption{Comparison among spectra of SN 1987A, SN 2007od, SN 2005cs, SN 1999em and SN 1992H during the nebular phase. For references see Sect.~\ref{sec:cfr} and Tab.~\ref{table:snc}.} 
\label{fig:cfr_3}
\end{figure}

The last comparison (Fig.~\ref{fig:cfr_3}) is made for the nebular
phase with \a\/ \citep{87a2}, SN 2005cs \citep{05cs2}, \em\/
\citep{99em} and the luminous SN 1992H \citep{92h}. The \Ha\/
profile of \od\/ differs from those of the other SNe.  The central
peak is blue-shifted and there is a boxy shoulder on the blue side
(see Sec.~\ref{sec:df}).  Also [CaII] is blue-shifted and shows an
asymmetric profile. Instead, [OI] $\lambda\lambda$ 6300, 6363
doublet, [FeII] at 7000\AA\/ and NaID are barely detectable preventing
a detailed analysis of the line profiles.

\subsection{Expansion velocity and temperatures}\label{sec:ev}

\begin{figure}
\includegraphics[width=\columnwidth]{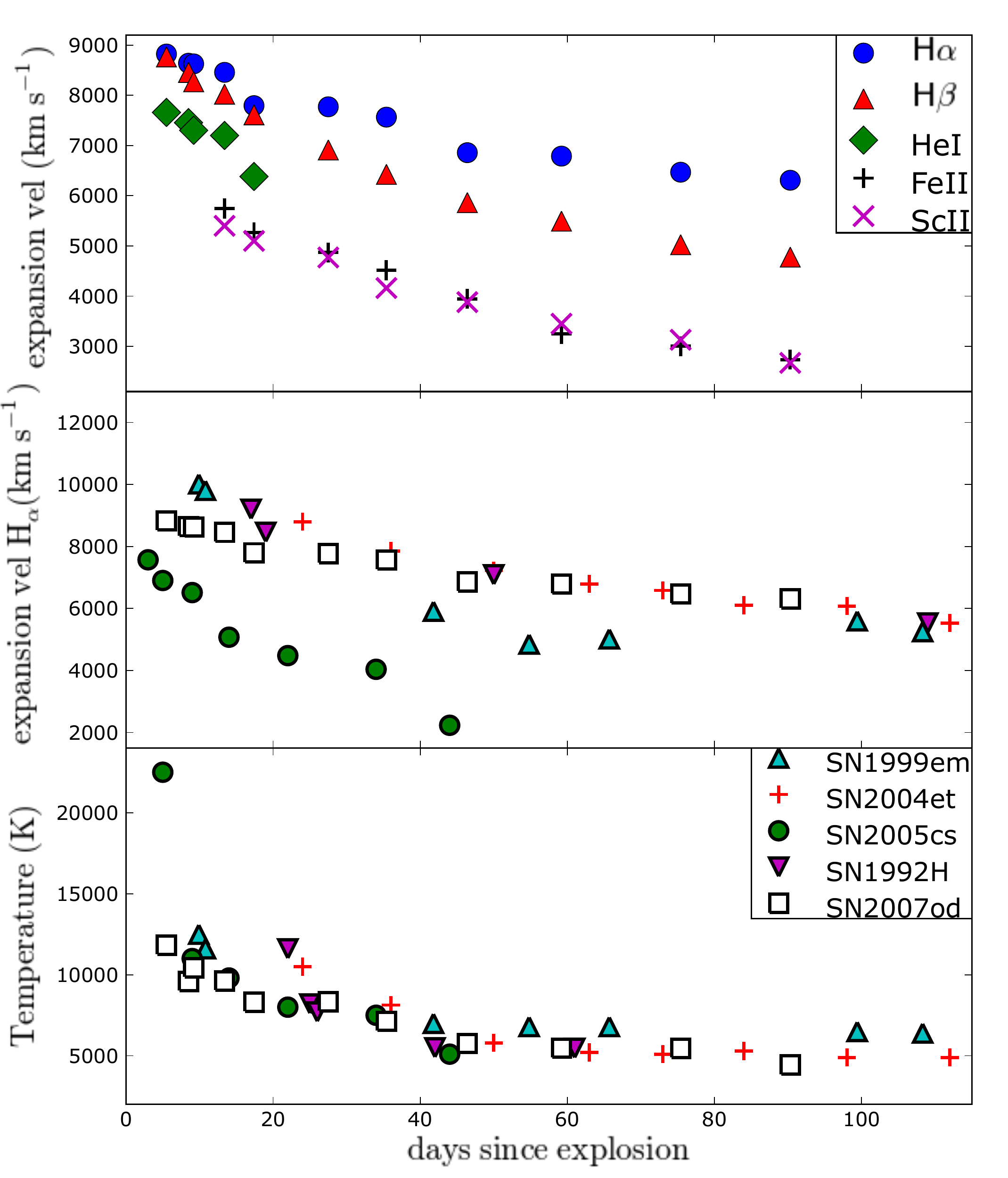}
\caption{Top: expansion velocity of \Ha, \Hb, He I $\lambda$5876, Fe II $\lambda$5169 and Sc II $\lambda$6246 as measured from the minima of the P-Cygni profiles. Middle: comparison of the \Ha\/ velocity of SN 2007od with those of other SNe II. Bottom: Evolution of the continuum temperatures T$_{bb}$ for the same SN sample.}
\label{fig:vel}
\end{figure}

The expansion velocities of \Ha, \Hb, He I 5876\AA\/,
Fe II 5169\AA\/ and Sc II 6246\AA\/, derived from fits to the
absorption minima, are reported in Tab.~\ref{table:op} and plotted in
Fig.~\ref{fig:vel} (Top panel).  Error estimates are derived from
the scatter of several independent measurements.
The velocities of \Ha\/ are comparable with those of \Hb\/ during the first 20
days, 
and progressively higher afterwards. During the 20 days in which the He
I line remains visible the velocity is about 1000 \kms\/ smaller
than that of \Ha.  Fe II velocity, which is a good
indicator for the photospheric velocity because of the small optical
depth, is lower than that of H and He I, and decreases below 3000
\kms\/ at about three months. Sc II is also a good indicator of the
photospheric velocity and its velocity is very close to that of Fe II,
supporting the identifications of both ions.

\begin{table*}
  \caption{Observed black-body temperatures and expansion velocities of \od.}
  \begin{center}
  \begin{tabular}{cclcccccc}
  \hline
  \hline
  JD & Phase$^{*}$ & T & v($H{\alpha}$) & v($H{\beta}$) & v(He I) & v(Fe II) & v(Sc II) \\
   +2400000 & (days) &(K) & (\kms)& (\kms)& (\kms)& (\kms)& (\kms)\\
 \hline
  54409.5 & 5.5 & 11830 $\pm$ 350 & 8822 $\pm$ 104  & 8764 $\pm$ 164& 7658 $\pm$ 122&& \\
  54412.5 & 8.5 & 9591 $\pm$ 350& 8639 $\pm$ 62& 8455  $\pm$ 140& 7454  $\pm$ 150&&\\
  54413.2 & 9.2 & 10429 $\pm$ 350& 8626 $\pm$ 60& 8270 $\pm$ 140& 7301$\pm$ 100 &&\\
  54417.4 & 13.4 & 9620 $\pm$ 350& 8457 $\pm$ 82& 8023 $\pm$ 150& 7199 $\pm$ 105& 5745 $\pm$ 132 &5400$\pm$ 200 \\
  54421.4 & 17.4 & 8305 $\pm$ 350 & 7794 $\pm$ 66 & 7610 $\pm$ 114 & 6382 $\pm$ 137 & 5270 $\pm$ 280 & 5100 $\pm$ 200\\
  54431.5 & 27.5 & 8328 $\pm$ 350 & 7771 $\pm$ 75 & 6912 $\pm$ 130 & & 4869 $\pm$ 200 &4770 $\pm$ 110\\
  54439.4 & 35.4 & 7123 $\pm$ 350& 7565 $\pm$ 75 & 6425 $\pm$ 400 & &4515 $\pm$ 110 & 4161$\pm$ 250 \\
  54450.4 & 46.4 & 5755 $\pm$ 350& 6857 $\pm$ 75& 5863 $\pm$ 133 & & 3946 $\pm$ 120 & 3880 $\pm$ 200  \\
  54463.2 & 59.2 & 5470 $\pm$ 350& 6788 $\pm$ 75& 5493 $\pm$ 128& &3250 $\pm$ 116 & 3450 $\pm$ 300\\
  54479.4 & 75.4 & 5450 $\pm$ 350 & 6468 $\pm$ 95& 5024 $\pm$ 100 & &3001$\pm$ 120 &3130 $\pm$ 200\\
  54494.3 & 90.3& 4429 $\pm$ 350& 6308 $\pm$ 101 & 4777 $\pm$ 102& & 2736 $\pm$ 110 & 2670 $\pm$ 100\\
 \hline
\end{tabular}
\begin{flushleft} 
$^{*}$ with respect to the explosion epoch (JD 2454404)
\end{flushleft} 
\end{center}
\label{table:op}
\end{table*}
  
In Fig.~\ref{fig:vel} (middle) we compare the \Ha\/ velocity
evolution of \od\/ with those of our comparison sample of Type IIP. In
the first months, the \Ha\/ velocity of \od\/ is comparable to
those of SN 1999em, SN 2004et, SN 1992H, and higher than that of \cs, which is
known to have slow photospheric expansion. As a major difference from
other objects the velocity decrement is always rather constant, even
in early phases.  The early velocity of \od\/, lower than those of
the other SNe of our sample, might be attributed to early-phase
interaction with a low density, thin CSM (cfr. Sect.~\ref{sec:cfr}).

The early spectra of \od\/ are fairly blue, suggesting moderately hot
black-body temperatures (T$_{bb}=1.1 \times 10^{4}$ K). In
Fig.~\ref{fig:vel} (bottom) the evolution of the temperature
derived from the blackbody fits to the spectral
continuum 
is shown and compared with those of the reference sample. The
temperature evolution of \od\/ is rather normal, although its
temperature during the very first days past
explosion 
never reached 12000 K.  About 40 days after explosion the temperature
becomes constant, in analogy to what observed in SNe 2004et, 1999em
and 1992H. This phase corresponds to the beginning of the H envelope
recombination.

\section{Dust formation and CSM interaction}\label{sec:df}
Already 40 years ago it was suggested that SNe could be an important
source of dust in the interstellar medium (ISM) \citep{ce,hw}. Recent
studies on the origin of dust \citep{tf,no,dw} have supported this
view, calling for core-collapse SNe as significant sources of dust in
the Universe. In fact, a number of objects have shown clear evidence of
dust formation, e.g., type IIn SNe 1998S \citep{98s,98s2}, 1995N
\citep{95n}, and 2005ip \citep{05ip}, type IIP SNe 1999em
\citep{99em}, 2004et \citep{04et,04et2}, 2004dj
  \citep{04dj2,04dj,04dj3} and 2007it \citep{07it}, type IIb SN
1993J \citep{95n}, and the peculiar SN 1987A \citep{87a1,87a4}, SN
2006jc \citep{06jc} and SN 2008S \citep{08s}.  Though these data
support the scenario of CC-SNe being important dust producers, the
amount of dust \citep[0.1--1.0 \M ,][] {dw,meikle} required to
explain the dust at high redshift 
is a few orders of magnitude larger than that measured in individual local
core-collapse SNe.

\begin{figure}
 \includegraphics[width=\columnwidth]{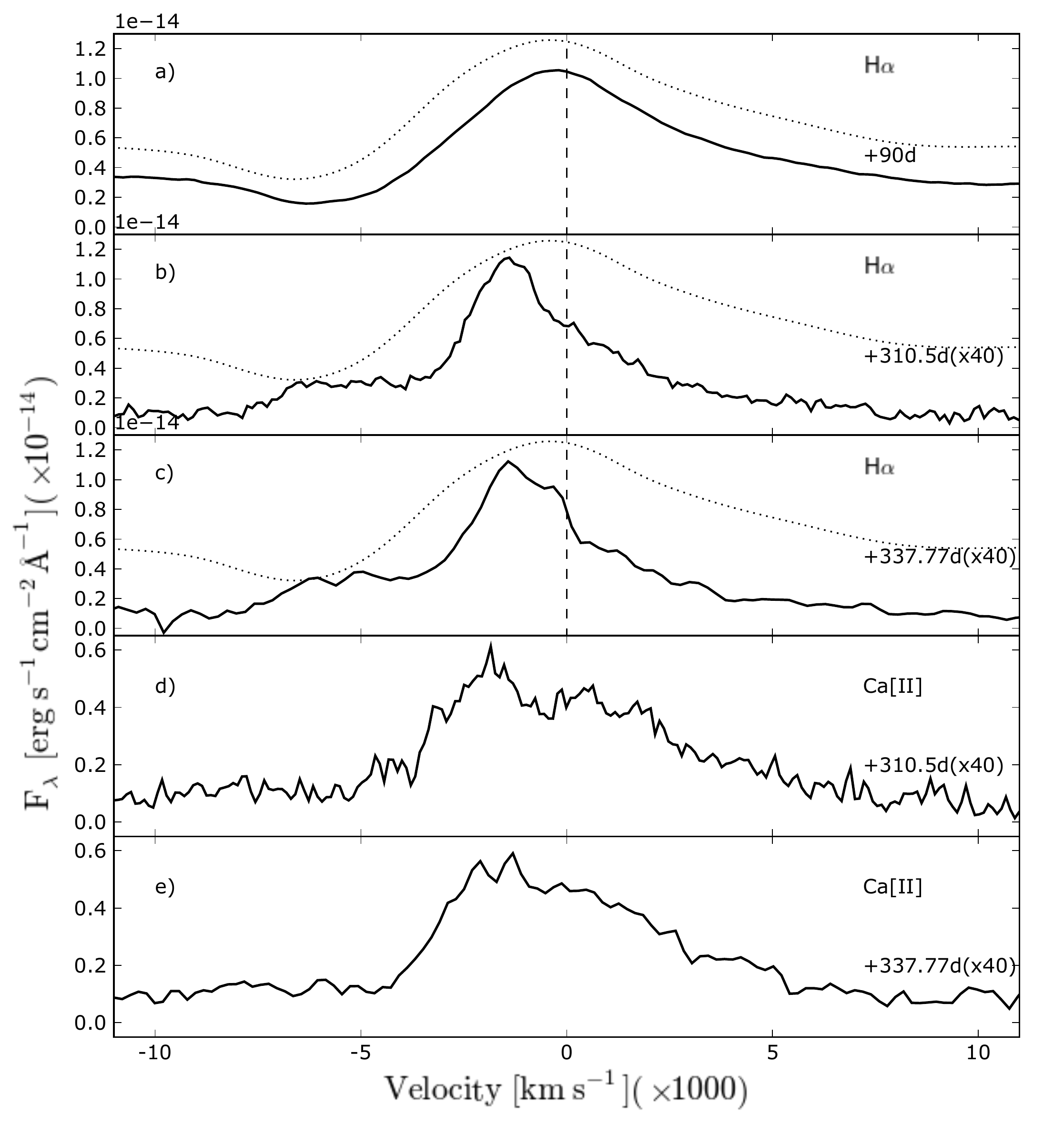}
 \caption{Zoom in the \Ha\/ region of the \od\/ spectra at 90d (panel a), 310d b) and 337d c).
The dotted line is the photospheric spectrum on day 60 used as a comparison. Panels d) and e) show the [CaII] profiles at late epochs.
The abscissa is in expansion velocity coordinates with respect to the rest frame positions of  \Ha\/ and to the average position of the [CaII] doublet. Phases relative to the explosion (JD 2454404) are indicated on the right. } 
 \label{fig:Han}
\end{figure}

\begin{table*}
  \caption{Evolution of the \Ha\/ line profile.}
  \begin{center}
  \begin{tabular}{lccccccc}
  \hline
  \hline
  Phase$^{*}$  & \multicolumn{5}{c}{Velocity$^{\diamond}$ (\kms\/)} & Flux$^{\dagger}$  \\
  (days) &  \multicolumn{5}{c}{---------------------------------------------------------------} &\\
         &  Blue  &  \multicolumn{3}{c}{Central Peak} &  Red & \\
         & edge &   position   & FWHM & HWHM$^{\ddagger}$  & Edge & \\
 \hline
 60 &  & -301 & & &  &\\
 75  &  & -319 & & &  &\\
 90  &  & -240 & & &  &\\
 226$^{**}$ & -8270 & -1325 & 4110 & 2100 & 7770 &51.20\\
 303$^{**}$ & -8000 &  -1500 & 2790 & 2010  &7860 &22.50\\
 310 & -8100 &  -1508 & 2815 & 2230 & 7700 &22.21\\
 337 & -8090 &  -1568 & 2623& 2280  & 7680&12.43\\
 342$^{**}$ & -8180 & -1508 & 2615 & 2200  & 7900 &14.30\\
 452$^{**}$ & -8100 & -1100 & 2970 &  2600  & 7850 & 5.28\\
 666/692$^{**}$ & -7400 &  & & & &\\ 
 \hline
 \end{tabular}
 \end{center}
 \begin{flushleft}
 $^{*}$ with respect to the explosion epoch (JD 2454404). \\
 $^{\diamond}$ with respect to the \Ha\/ rest frame wavelength. \\
 $^{\dagger}$ 10$^{-16}$ erg s$^{-1}$ cm$^{-2}$\\
 $^{\ddagger}$ HWHM of the blue, unabsorbed wing computed with respect to the rest frame wavelength.\\
 $^{**}$ spectra from \citet{07od}, flux calibrated with broad band photometry. 
 \end{flushleft}
\label{table:Haneb}
\end{table*}

There are manifold observational signatures of the formation of dust
in SN ejecta.  Examples are the dimming of the red wings of
line profiles due to the attenuation of the emission originating in
the receding layers, and the steepening of the optical light curves.
These effects were observed in \a\/ \citep{87a1,87a4} , SN~1998S \citep{98s}, 
\em\/ \citep{99em} and \et\/ \citep{04et}.
Additional evidence is a strong IR excess observed in a few objects at
late times, e.g. SNe 1998S and 2004et, but also, though very rarely,
at early times, as in SN~2006jc.  We stress that for the comprehension
of the physical processes occurring in the aftermath of the
explosions it is important to distinguish between IR thermal emission
from newly formed dust within the SN ejecta and IR echoes of the
maximum-light emission by pre-existing, circumstellar material (CSM).

A distinguishing precursor of dust formation is the detection of
rotation-vibration molecular lines of CO, which are powerful coolants.
Indeed CO was observed in several SNe, e.g., SNe 1995ad
\citep{95ad}, 1998dl and 1999em \citep{98dl}, 2002hh \citep{02hh},
2004dj \citep{04dj2} and 2004et \citep{04et,04et2}, in which
also dust formation in the ejecta was detected. The presence of
CO molecular lines seems, therefore, a necessary condition for dust
condensation in the ejecta, as reported by \citet{98s}.

Another important issue is the site of dust formation in SNe. Indeed
dust in core-collapse SNe has been detected: a) deep within the
ejecta, e.g., in SNe 1999em and 1987A, or b) in a cool dense shell
(CDS) created by the SN ejecta/CSM interaction, e.g. in SNe 1998S,
{2004dj and 2004et.  The scenarios are not exclusive and evidence
of both phenomena was found, as shown in \citet{04et2}.
Unfortunately, the paucity of MIR late time observations makes the
determination of the site of dust formation very difficult.

The signature of dust formation in \od\/ is provided by the blue-shift
($\sim$ 1500 \kms\/) of the peak of \Ha\/ and by the
corresponding attenuation of the red wing, which is seen in late time
spectra (see the panels of Fig.~\ref{fig:Han}).  On the other
hand, the decay rates of the optical light curve between 208d and 434d
(including also data by \citet{07od}, cfr. Tab.~\ref{table:mdata}) are
in close agreement with the decay of $^{56}$Co, suggesting that the
dust formed during the period of un-observability (February to May
2008), i.e. before the late-time SN recovery (cfr.
Sect.~\ref{sec:pe}).  Indeed \citet{07od} show that the optical
and IR SED of \od\/ at about 300d can be fitted by the sum of a
black-body originating in the SN ejecta plus a cooler black-body
emission at 580 K due to dust.

In addition to the \Ha\/ skewed central peak, the line profile
(Fig.~\ref{fig:Han}) shows also the presence of structures.  A boxy
blue shoulder extends to about --8000 \kms (cfr.
Tab~\ref{table:Haneb}), then drops rapidly to zero.  This is
reminiscent of the boxy profiles of late-time interacting SNe II SNe
1979C, 1980K \citep[][ and references therein]{fesen99}, 1986E
\citep{86e} which have been interpreted as evidence of interaction of
the SN ejecta with a spherical shell of CSM \citep{chev3,chev}.  We
note that, unlike other SNe, in \od\/ a red, flat shoulder is
not visible, possibly because it is attenuated by dust intervening
along the line of sight.

A hint of a narrow, unresolved emission at the \Ha\/ rest
wavelength, more evident in the Gemini spectra by \citet{07od}, is
visible also in our latest spectrum.  The peak of the [Ca II]
$\lambda\lambda$ 7292, 7324 emission is also blue-skewed with
attenuation of the red wing in analogy to \Ha\/ (panels d and e
of Fig.~\ref{fig:Han}). Instead, [Ca II] does not show the blue
shoulder indicating that the interaction does not affect the
metal-enriched ejecta, as noted in \citet{07od}.

\citet{07od} noticed analogies between the \Ha\/ profiles of
\od\/ and \s, and tried to explain the \Ha\/ profile
between 8 and 20 months as the combination of different phenomena: the
interaction of the ejecta with a CSM torus and a blob of CSM out of
the plane of the torus, plus the possible presence of a light echo.
Each process corresponds to specific components of the line profile
(cfr. their Fig.~4): two components at about $\pm 1500$ \kms, that arise
in the radiative forward shock of the ejecta interacting with the
torus which should be highly inclined to produce the low observed
projected velocities; a slightly broader component at about $-5000$
\kms\/ due to the ejecta--blob interaction with a larger velocity
component along the line of sight; a residual ejecta component similar
to that observed on day 50, required to explain the broad red wing of
the line; and a light echo that became dominant about 2 yr after
explosion.  The blue-shifted component at $-1500$ \kms\/ appears
stronger than that at $+1500$ \kms.  This asymmetry was attributed to
strong extinction suffered by the receding component due to dust
formed in the CDS.  The weakness of this scenario is in the geometry
of the torus.  As mentioned above, the torus is required to have an
inclination of the order of $80 \degr$ (for H expansion velocity of
the order of 10000 \kms) to explain the small projected velocities.
Even for a thick torus, it is difficult to reconcile the required high
inclination with the strong extinction suffered by the red component.
For the same ejecta velocity the inclination of the blob along the
line of sight is about $60\degr$, i.e. it is not orthogonal to the torus.

\begin{figure}
 \includegraphics[width=\columnwidth]{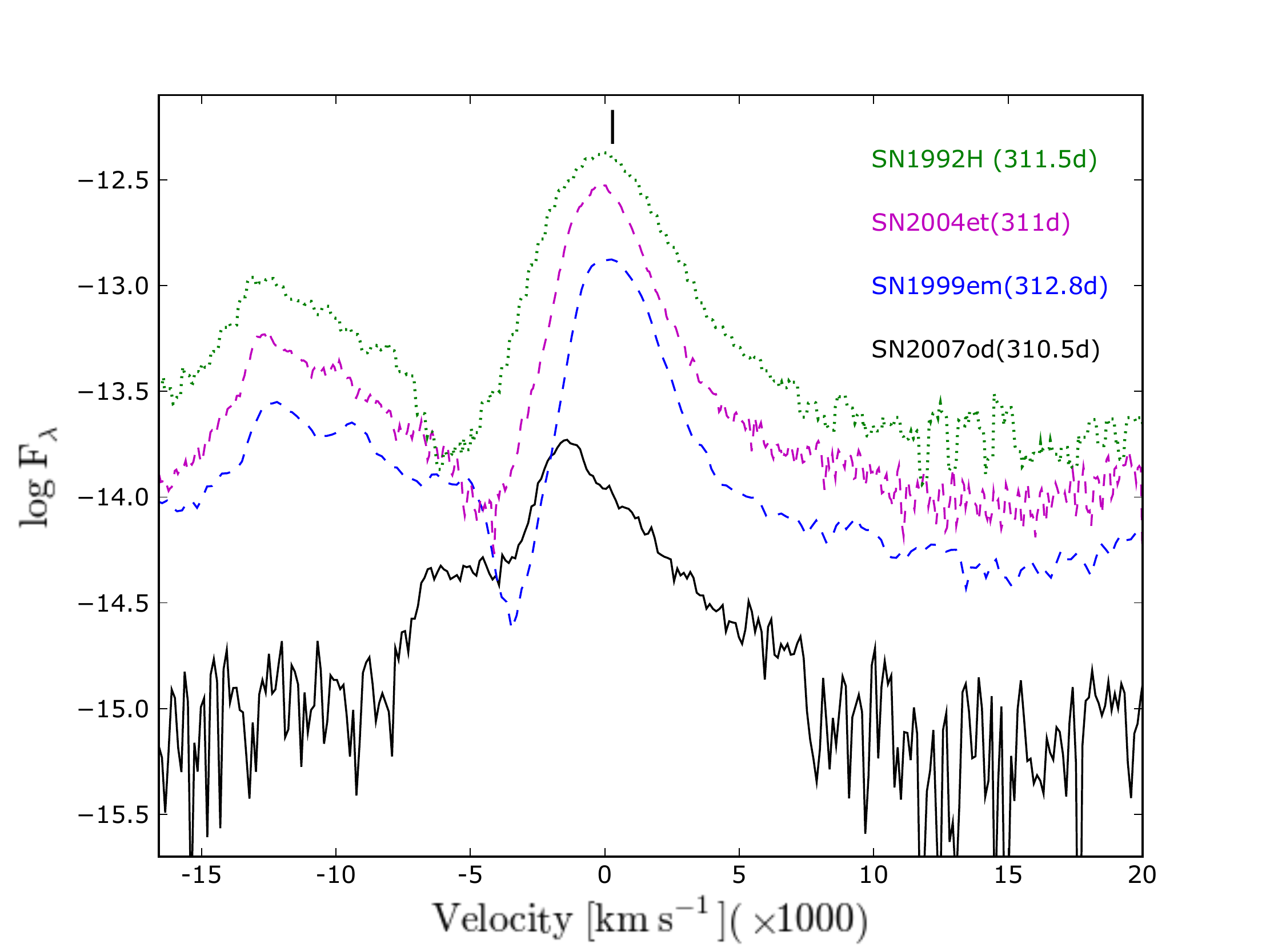}
 \caption{Comparison of the  \Ha\/ profile of \od\/ during the nebular phase with those of SNe 1992H, 1999em and 2004et. The spectra of all SNe were reported to the same distance of \od. The position of the \Ha\/ rest wavelength is marked with a vertical dash.} 
 \label{fig:spdist}
\end{figure}

Useful information on the phenomena taking place in \od\/ at late
times come from the comparison of the late \Ha\/ profile with
those of other SNe IIP. In Fig.~\ref{fig:spdist} all spectra were
corrected for absorption and rescaled to the distance of \od. The line
flux of \od\/ is significantly smaller than those of other SNe.  The
blue, (likely) unabsorbed wing of the central/core \Ha\/
emission of \od\/ roughly coincides in wavelength with the
corresponding blue wings of other non-interacting SNe II at similar
phases (we stress that SN 1999em at this epoch does not show sign of interaction).  
In particular, the HWHM(\Ha)$\sim2000$ \kms\/
(Tab.~\ref{table:Haneb}) of the blue side is compatible with the
FWHM(\Ha)$\sim$4000 \kms\/ of the emission of \h\/ at 311
days after explosion, a SN that has a similar kinematics to \od\/ (cfr.
Fig.~\ref{fig:vel}). Also the terminal velocities of \Ha\/ seem
very similar, since the edge of the blue shoulder of \od\/ extends out
to 8000 \kms, a velocity comparable to the bluest wing of the residual
P-Cyg absorption of \h\/.  The flux of the blue wing of the central
emission (at about 6500\AA), likely less absorbed, coincides with
those of other SNe II, e.g. \em.  On the contrary, the flux at the
rest wavelength is significantly depressed, and even more is the red
wing, indicating the presence of dust within the ejecta. The resulting
profile is skewed.  The comparison, therefore, seems to point towards a
common origin of the line cores extending between 6450 and 6650\AA\/
as arising from the spherical expanding ejecta.

Evidence of dust in the ejecta was also seen in \a\/ and \em.  In the
case of SN 1987A the dust formed in an inner core at v$\sim$1800
\kms\/ with optical depth $\tau \leq 1$ at a much later epoch
t$\sim$526d \citep{lucy} than in \od\/. Also in \em\/ the dust formed
late (t$\sim$500d) when no sign of interaction was present. Here the
dust location, within an inner region at v$\sim$800 \kms\/ with $\tau
\geq 10$, was derived from a careful analysis of the profile of the [O
I] $\lambda\lambda$6300-6363 doublet \citep{99em}.  As already
mentioned, the [O I] doublet is barely visible in the late spectra of
\od\/ and the analysis relies on the line profile of \Ha\/
which arises mainly from the outer ejecta.

As for the CSM/ejecta geometry proposed by \citet{07od} to explain the
\Ha\/ profile, the formation of dust deep within the ejecta of
\od\/ is also not devoid of problems. In fact, at the early epoch of
occurrence (t$\leq226$d) the SN ejecta are expected to be too warm for
dust condensation \citep{kozasa}, while the presence of interaction at
this phase justifies the dust formation in an outer CDS.  
It is possible that \od\/ was intrinsically less energetic
than average CC-SNe, similar to low--luminosity SNe like 1997D and
2005cs
and that the temperature at about 200 days was already low enough to
allow dust condensation in some regions of the SN ejecta.

Observations show first evidence of interaction already on day 226. At
this epoch the ejecta reached the material expelled by the
progenitor in mass loss episodes at short times before explosion.  The
interaction produces forward and reverse shocks \citep{chev3}. In
general, between the two shocks the gas undergoes  thermal
instability and cooling, thus creating a cool dense shell (CDS) in which
subsequently dust can form. The formation of a CDS can take place
behind the forward shock primarily in the CSM, as invoked in the case
of SN 2006jc by \citet{06jc}, or behind the reverse shock as in \s\/
\citep{98s}. With a standard mass loss rate, the CDS forms behind the
reverse shock, i.e. in the ejecta that are denser and chemically
richer, hence more prone to dust formation \citep{98s}.  The presence
of dense clumps in the CSM, which eventually are overcome by the fast
SN ejecta, may allow the formation of a CDS (and dust) also in regions
that eventually are deeply embedded in the ejecta.

\begin{figure}
 \includegraphics[width=\columnwidth]{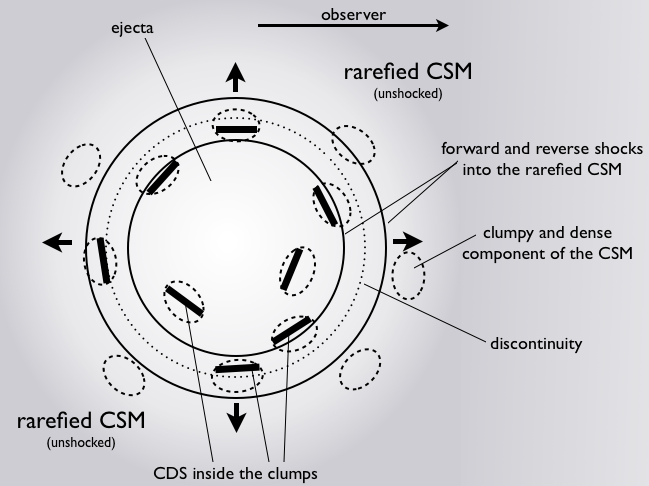}
 \caption{Schematic illustration of the geometry of the newly formed dust in \od. The CDS arise both inside the dense clumps and between the reverse shock and the discontinuity of the rarefied component of the CSM.} 
 \label{fig:cds}
\end{figure}

We believe that the observed late-time line profiles at about 300 days
can be explained schematically with: 1) a relatively broad skewed
emission from SN ejecta (FWHM$\sim$2800 \kms) typical of SNe II at
corresponding epochs, significantly absorbed by dust formed {\it
  within the ejecta, possibly in clumps}; 2) a distorted boxy profile
with high velocity (up to 8000 \kms), evidence of the interaction of
the H-rich ejecta with a {\it spherical} outer CSM, with the red
shoulder depressed due to dust absorption. The presence of late time
interaction is supported by the re-brightening of optical light curves
by day $\sim$653, as shown by \citet{07od}; 3) an unresolved emission
which forms in the un-shocked CSM \citep[cfr. ][]{07od}.

A plausible scenario to explain the observed structure of \Ha\/
could be similar to that proposed for SN 1988Z \citep{88z}. The ejecta
collides with a two-component wind: a spherically symmetric substrate
composed of a relatively rarefied gas, and very dense clumps (cfr.
Fig.~\ref{fig:cds}). Such structure allows the formation of the boxy
profile due to the interaction between the fast expanding ejecta and
the CSM and, at the same time, the early formation of dust deep in
the ejecta in the CDS of the inner clumps.  The dust inside the ejecta
progressively screens the radiation arising from the receding layers
and skews both the central profile from the unperturbed ejecta and the
boxy profile from the interaction.  Indeed the profile of \Ha\/
suggests the presence of dust also in the inner ejecta since dust
formation in the outer CDS dims both the blue and red
wings in the same way, and the red shoulder would appear boxy and not tilted
(Fig.~\ref{fig:Han}).
 
The analysis of the IR SED performed by \citet{07od} shows a strong IR
excess attributable to warm dust, consistent with an amorphous-carbon
dominated model with 75$\%$ amorphous-carbon and 25\% silicate.
The radiative transfer models used by \citet{07od} suggest up to 4.2 x
10$^{-4}$ \M\/ dust, mainly composed of amorphous-carbon grains
formed in a CDS.
 
The flat blue shoulder of \Ha\/ is barely visible in the
spectrum taken on day 226 but is very well formed on day $\sim$337
when the emission extends out to $\sim$8200 \kms, and remains well
defined until day 680 \citep{07od}.  Assuming that the CSM is due to
the mass loss of material traveling at 10 \kms, this indicates the
presence of CSM in a range of distances of about 1300--1950 AU and that
the progenitor experienced enhanced mass loss 500--1000 yr before the
explosion.

\section{Discussion}\label{sec:dis}
In the previous sections we presented new data of the type II \od\/ in UGC~12846,
including for the first time photometric and spectroscopic
observations of the early epochs after the explosion.

Analysis of the multicolor and bolometric light curves indicates that
at early times \od\/ was a luminous SN IIP like SNe 1992H and 1992am.
The absolute magnitude at maximum, M$_{V}=-18.0$, the bolometric
luminosity L$_{bol}=6.0 \times 10^{42}$ erg s$^{-1}$ (see
Tab.~\ref{table:mdata}), and the relatively short plateaus in V and R
suggest an envelope mass smaller than that of standard plateau events.

\begin{table}
\caption{Main data  of SN~2007od}
\begin{center}\tabcolsep=0.95mm
\begin{tabular}{lcc}
\hline
position (2000.0)	&23$^h$55$^m$48$^s$.68	&+18$^o$24$^m$54$^s$.8 \\
parent galaxy			& \multicolumn{2}{c}{UGC~12846,  Sm:}\\
offset wrt  nucleus 		& 38$\arcsec$E		& 31$\arcsec$S 	\\
adopted distance modulus	&  $\mu=32.05\pm0.15$	&	\\
SN heliocentric velocity	&  $1734\pm3$ \kms	&	\\
adopted reddening		& E$_{g}$(B-V)$=0.038$	& E$_{tot}$(B-V)$=0.038$\\
\hline
\end{tabular}
\begin{tabular}{lccc}
	& peak time		& peak observed	& peak absolute  \\
	&  (JD 2454000+)	&  magnitude		&  magnitude	\\
U	& $409\pm2$		&$13.490\pm0.04$		&$-18.7\pm 0.18$	\\
B	& $409\pm2$		&$14.464\pm0.03$		&$-17.8\pm 0.18$\\
V	& $411\pm2$		&$14.144\pm0.02$		&$-18.0\pm 0.23$\\
R	& $411 \pm2$		&$13.945\pm0.02$ 	&$-18.1\pm 0.21$ \\
I	& $411\pm2$		&$13.821\pm0.02$	&$-18.2\pm 0.23$\\
uvoir	& $410\pm2$	& \multicolumn{2}{c}{L$_{bol}=6.0 \times 10^{42}$ erg s$^{-1}$}  \\
	&				&				&				\\
rise to R max & $\sim5$ days	& 			&			\\
explosion day & $\sim404\pm5$ & \multicolumn{2}{c}{$\sim30$ Oct. 2007}			\\
\hline
\end{tabular}
\begin{tabular}{p{4.4cm}cc}
	& late time decline	&interval \\
	& mag(100d)$^{-1}$		& days\\
V	& 	0.94			& 208--434\\
R	& 	1.00			& 208--434\\
I	& 	1.08			& 208--434\\
uvoir&       1.053			& 208--434\\
UBVRI&    1.065		& 208-434\\
\hline
\end{tabular}
\begin{tabular}{p{6cm}c}
M(Ni)			&   0.02 \M \\
M(ejecta)			& 5-7.5 \M \\
explosion energy	&  0.5$\times 10^{51}$  ergs\\
\hline
\end{tabular}
\end{center}
\label{table:mdata}
\end{table}

\od\/ shows several interesting properties.  At early epochs
(t$< 15$d), T$_{bb}$ and v$_{exp}$ are lower than in other bright SNe
II, e.g. \h\/ and \et.  In the same period the \Ha\/ emission
component is squared, and there is evidence for HV features in
\Hb\/ and \Hg.  These properties, together with 
bright luminosity and shallow absorption features, point towards early interaction of the ejecta with a
low density CSM \citep[cfr.][]{moriya}.  
The absence of X-ray or radio detection is not against this scenario, because the distance to SN 2007od is 26 Mpc, much larger than that of SN 1999em (7.5-7.8 Mpc). 
If for SN 2007od we assume  the same radio and X-ray luminosity of SN 1999em, the observed flux should be below the treshold of the observation. Indeed the X-ray flux of SN 1999em was close to the limit (about 10$^{-14}$ ergs cm$^{-2}$ s$^{-1}$) as well as the radio flux reported in \citet[][]{pooley}. 
If SN 2007d had the same emission of SN 1999em, the expected X-ray flux would be about 10$^{-15}$ ergs cm$^{-2}$ s$^{-1}$, much lower
than the 3$\sigma$ upper limit by \citet{a1}. Moreover, HV features in optical spectra were proposed by \citet{chugai}, as clues of interaction and detected in SN 1999em and SN 2004dj.  We claim that this is the case also for SN 2007od.
A rough estimate of the CSM
mass in close proximity to the progenitor can be derived starting by SYNOW parametrization of the
early spectra.  Through the optical depth of high-velocity lines ($\tau$=0.8) we can gain the H density in the transition.
Thanks to the Saha equation in LTE approximation it is possible to obtain the ratio between HI (n$_{HI}$=1.7$\times$10$^{-30}$ -- 1.4$\times$10$^{-31}$ based on different assumptions) and the total H density. This ratio is roughly related to the mass of the CSM. In our case we estimated a CSM mass of the order of $\sim$ 10$^{-3}$ -- 10$^{-4}$ M$_{\odot}$. This value is in agreement with
the velocities of the HV lines reported in Sec.~\ref{sec:sa}, in fact
a greater amount of CSM would not allow reaching such velocities.

The light curves show a drop of $\sim$6 mag from the plateau to the
tail.  The tail is therefore relatively faint, corresponding to a \ni\/ mass
M(\ni)$\sim$ 3x10$^{-3}$ \M, unusually small for SNe~IIP that are
luminous at maximum and are comparable to those estimated in faint
SNe IIP such as SN~2005cs.  In Sect.~\ref{sec:df} we show  
spectroscopic evidence of dust formation at late times and
therefore that the derived amount of \ni\/ should be considered as a
lower limit.  A more solid estimate, based on the bolometric flux
including the MIR late time emission by \citet{07od}, is
M($^{56}$Ni)$\sim$ 2x10$^{-2}$ \M, thus indicating that about 90\% of
the optical+NIR emission is reprocessed by dust.

Dust formation occurs within day 226 after the explosion, quite early
in comparison to other core-collapse SNe, e.g. \a\/, \em\/ and SN
2004et. To our knowledge, \od\/ is the type IIP SN showing the
earliest dust formation.  Late-time optical spectroscopy shows also
clear signs of strong ejecta--CSM interaction. Indeed, the complex
\Ha\/ profile can be interpreted as the combination of a
typical ejecta emission, ejecta--CSM interaction, and the presence of
dust in clumps.  However, the combined effect of interaction and dust
does not affect the decline rate that is similar to that of
$^{56}$Co decay from 208 to $\sim$434 days past explosion
(Fig.~\ref{fig:07_bol}).
CSM-ejecta interaction makes a minor contribution to the bolometric
luminosity and only affects line profiles.  A flattening of the light
curve might be present after day 600 as shown in \citet{07od},
either because of increased interaction or a light echo, but the data
are too inhomogeneous for a strong statement in this sense.

Because of the location in the outskirts of the parent galaxy, the
metallicity at the position of \od\/ could not be determined from our
spectra nor were we able to find spectra of the host galaxy in the main
public archives.  UGC 12846 is classified as Magellanic Spiral (Sm:)
of low surface brightness (LSB), as confirmed by our
Fig.~\ref{fig:sn07od} (left) in which the host galaxy is barely
visible.  LSB galaxies are objects with peak surface brightness
$\mu_{peak}^B\geq$ 22.5-23 mag arcsec$^{-2}$ having generally low
metallicity (Z$<\frac{1}{3}$Z$_{\odot}$ \citet{mcgaugh};
0.1$<$Z(LSB)/Z$_{\odot}$$<$0.5, with no radial dependence
\citet{bev}).  \citet{smo1} and \citet{smo} for UGC~12846 report a
central surface brightness $\mu_{peak}^{B}\sim$22.65 decreasing to
$\mu>26$ mag arcsec$^{-2}$ at position of the SN. Using $M_{B}=
-16.79$ \citep{smo1} and the diagram shown in \citet{mcgaugh} we can
consider UGC 12846 as a LSB with average oxygen abundance. Therefore,
the environment of \od\/ is likely metal poor with log(O/H) $\sim-4$.
In general, we expect that low metallicity stars suffer little mass
loss and have big He cores and massive H envelopes when they die
\citep{heger}.  The significant presence of CSM both at early and late
times, a clear evidence of mass loss, seems in disagreement with this
scenario, although recently \citet{chev2} suggested that there are
mass loss mechanisms that do not decline at lower metallicities.
Alternatively, the presence of companions can explain strong mass
loss by metal poor stars.

In nebular spectra of core-collapse SNe the flux ratio $R= \frac{[Ca
  II] \lambda\lambda7291,7324}{[O I] \lambda\lambda6300,6364}$ is
almost constant with time \citep{frans,frans1}. This ratio, only
marginally affected by differential reddening, is a useful diagnostic
for the mass of the core and consequently for estimating the
progenitor mass, with small ratios corresponding to higher main
sequence masses. In SN 1987A it was $R\sim$3 \citep{elm}, for the
faint SN 2005cs it was $R\sim$4.2$\pm$0.6 \citep{05cs2}, while for SNe
1992H and 1999em we computed the values $R\sim$1.61 and
\textit{R}$\sim$4.7, respectively.  In comparison, the measured ratio
$R\sim$32$\pm$5 of \od\/ is very high, suggesting that the progenitor
mass is quite small.

A super-asymptotic giant branch (hereafter SAGB) progenitor with a
strongly degenerate Ne-O core might explain these observables. As
shown by \citet{pumo}, the most massive SAGB stars (M
$\lesssim$ M$_{mas}$ $\sim$ 10 -- 11 M$_{\odot}$) can indeed suffer strong
episodes of mass loss while still preserving significant H envelopes
($\gtrsim$ 5-9 M$_{\odot}$) at the end of their
evolution. 
In fact, the outcome of SNe from a super-AGB progenitor can differ
according to the configuration of the super-AGB star at the moment of the
explosion, and may range from a type II SN (either IIP or IIL
depending on the mass of the envelope) with relatively low degree of
CSM interaction, to a type IIb SN having stronger interaction with the
CSM, up to a stripped-envelope SN.  The high value of \textit{R} could
be explained also with Fe core collapse SN II in the low limit of the
mass range (M$_{\odot}\sim$11-12); these stars might produce
oxygen-poor SN II.  But these stars hardly explain the episodes of
mass loss at metallicities lower than solar \citep{woosley2}.  Binary
companions could easily explain the episodes of mass loss, regardless
of the progenitor star.  We cannot exclude {\it a priori} the binary
system solution.

The formation of dust plays a key role in shaping the display of SNe
II.  In SN 1998S, \citet{98s} calculated a value of $\sim$10$^{-3}$
\M\/ of dust in a CDS. \citet {99em} obtained a lower limit of
$\sim$10$^{-4}$\M\/ in the ejecta of \em\/ from the analysis of the [O
I] 6300 \AA\/ evolution. For \a\/ \citet{87a3} reported a dust mass of
7.5$\times$10$^{-4}$ \M. For SN 2004et \citet{04et2} estimated a dust
mass of M$=(2-5)\times$10$^{-4}$ \M\/ in the CDS, and through
hydrodynamical considerations suggested that the mass of new dust
produced either in the ejecta or in a CDS never exceeded 10$^{-3}$
\M.  The formation of 1.7 to $4.2\times10^{-4}$\M\/ of dust
\citep{07od}, contributes to the unusual drop of $\sim6$ mag (from
plateau to tail) in \od.

The dust mass estimate can change by an order of magnitude if we
consider silicates or carbon grains and if we take into account the
possibility of very opaque clouds \citep{98s,99em}.  Indeed, when
forming in clumps most of the dust could be undetectable due to the
clump opacities, and be in much higher amounts than necessary to
produce the observed blackbody emission.  We note that the environment
of this SN is different from other dust forming SNe that normally
explode in spiral galaxies or in regions with solar metallicity.
Indeed, the environment metallicity of \od\/ is between solar and that
of galaxies at \textit{z}$\gtrsim$6. Thus, one may speculate that
larger amounts of dust are formed at high redshift because of low
metallicities.

\subsection{Explosion and progenitor parameters}\label{sec:m}

We estimated the physical properties of the SN progenitor
(namely the ejected mass, the progenitor radius, the explosion
energy) by performing a simultaneous $\chi^{2}$ fit of the main
observables (i.e., bolometric light curve, evolution of line
velocities, and continuum temperature at the photosphere) against
model calculations, in analogy to the procedure adopted for other SNe
\citep[e.g., SNe 1997D, 1999br, 2005cs,][]{zamp2,zamp3}.

Two codes were used to produce models: a semi-analytic code
\citep[described in detail in][]{zamp2} which solves the energy
balance equation for a spherically symmetric, homologously expanding
envelope of constant density, and a new, relativistic,
radiation-hydrodynamics code (described in detail in \citealt{pumo2}
and Pumo \& Zampieri, submitted). The latter is able to compute the
parameters of the ejecta and the emitted luminosity up to the nebular
stage by solving the equations of relativistic radiation hydrodynamics
in spherically symmetry for a self-gravitating fluid which interacts
with radiation, taking into account the heating due to decays
of radioactive isotopes synthesized in the SN explosion.

The semi-analytic code was used to perform a preparatory study in
order to constrain the parameter space and, consequently, to guide the
more realistic but time-consuming simulations performed with the
relativistic radiation-hydrodynamics code.  We note that modelling
with these two codes is appropriate if the emission from the supernova
is dominated by the expanding ejecta. For SN 2007od the contamination
from interaction and the formation of dust may in part affect the
observed properties of the supernova during the first $\sim 20$ days
and from the end of the plateau afterwards. However, there is no
evidence that either interaction or dust formation are important during
most of the photospheric phase ($\sim 20-90$ days) and hence our
hydrodynamic modeling can be safely applied to SN 2007od during this
time frame. This is more than sufficient for providing a reliable
estimate of the main physical parameters of the ejecta, with the
exception of the Ni mass that, as already mentioned, is inferred from
late time optical and NIR observations.

The shock breakout epoch (JD $= 2454404\pm5$) and distance modulus
($\mu= 32.05\pm0.15$) adopted in this paper (cfr. Sect.~\ref{sec:red})
are used to fix the explosion epoch and to compute the bolometric
luminosity of \od\/ for comparison with model calculations. Assuming a
$^{56}$Ni mass of 0.02 M$_{\odot}$ (see Sect.~\ref{sec:bol}), the best
fits of the semi-analytic and numerical models are in fair agreement
and return values of total (kinetic plus thermal) energy of $\sim 0.5$
foe, initial radius of $4-7 \times 10^{13}$ cm, and envelope mass of
$5-7.5$ \M\/ (Fig.~\ref{fig:model}).

\begin{figure}
 \includegraphics[width=9cm]{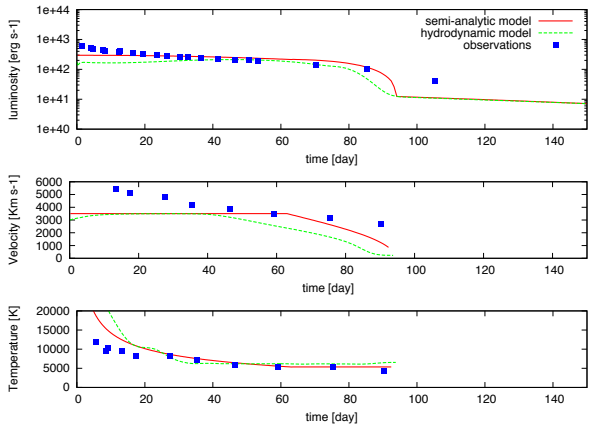}
 \caption{Comparison of the evolution of the main observables of SN
   2007od with the best-fit models computed with the semi-analytic
   code (total energy $\sim 0.5$ foe, initial radius $4 \times
   10^{13}$ cm, envelope mass $5$ \M) and with the relativistic,
   radiation-hydrodynamics code (total energy $\sim 0.5$ foe, initial
   radius $7 \times 10^{13}$ cm, envelope mass $7.6$ \M).  Top,
   middle, and bottom panels show the bolometric light curve, the
   photospheric velocity, and the photospheric temperature as a
   function of time respectively.  To estimate the photosphere
   velocity from observations, we used the value inferred from the Sc II
   lines (often considered in Type II SNe good tracer of the
   photosphere velocity).  }
 \label{fig:model}
\end{figure}

The best-fit models show some difficulty in reproducing the early part
of the light curve ($\lesssim$ 25 days) and the very end of the
apparent plateau (at $\sim$ 106 day).  The excess at early epochs may
be caused in part by the ejecta-CSM interaction (see discussion in
Sect.~\ref{sec:dis}).  As for the point at $\sim$106 days, the excess
may be related to the final interaction stage with the H-rich CSM that
occurs at the end of the recombination phase.

In Fig.~\ref{fig:model} we show also the evolution of the photospheric
velocity and temperature.  The agreement between our modeling and the
observational data is good apart from, once again, the early phase.
The reason for this difference may be due to both interaction and the
approximate initial density profile used in our simulations, which may
not reproduce correctly the radial profile in the outermost
high-velocity shells of the ejecta formed after shock breakout
\citep[e.g.,][Pumo \& Zampieri, in prep.]{utrobin}. For this reason, we
did not include the first 3 measurements of the line velocity in the
fit.

The values of the modeling reported above are consistent with the
explosion and mass loss of a SAGB star with an initial (ZAMS) mass
$\lesssim$ M$_{mas}$ \citep[M$_{mas}\sim 9.7$ to 11.0 \M\/ for the
environment metallicity limits of][cfr. Sect.~\ref{sec:dis}, see also
\citet{pumo} for details]{pumo}.  The moderate ejecta mass and amount
of \ni, the relatively low explosion energy, the very large Ca/O
ratio, and the presence of C rich dust fit reasonably well within this
framework.

Some of these constraints may appear consistent with a fallback
supernova from a massive star. In this scenario the mass-cut is
located sufficiently far out to trap part of the \ni\/ and
intermediate mass elements (e.g. O) that fall back onto the
compact object. Also the relatively low explosion energy may be
related to this mechanism \citep{fryer,zamp}. However, significant
mass loss prior to explosion is not easy to reconcile with a high mass
progenitor in a low metallicity environment (see Sect.~\ref{sec:dis}).
Problematic are also the moderate amount of \ni\/ estimated including
MIR data, the relatively high expansion velocity and low ejecta mass,
and the C-rich dust \citep{07od}, since usually these stars produce
silicate-rich dust \citep[][and reference therein]{04et2}.

\section{Conclusions}\label{sec:final}

In this paper we present new early-- and late--time observations of
\od.  The SN is among the brightest
type IIP SNe known to date (peak magnitudes M$_{R}=-18.1$ and
M$_{V}=-18.0$) and shows evidence of interaction with CSM both at
early and late times. These observations, along with dust formation at
late epochs, make this object different from common IIP events.

The light curve of \od\/ reaches a bright, short plateau lasting
$\sim$25d at M$_{R}=-17.8$ after a short post-peak decline in which
the luminosity decreases by 0.3 mag in a few days. The early hiigh
brightness is coupled with a low luminosity tail comparable to that of
the faint SNe, e.g. SN 2005cs. The magnitude drop from the plateau to
the first available point of the tail is $\sim$6 mag, unusually large
for a SN IIP. The bolometric light curve tail roughly follows the
slope expected for $^{56}$Co decay into $^{56}$Fe.  The $^{56}$Ni
mass derived from the late-time $uvoir$ (UV to K) luminosity is
M(Ni)$\sim$3x10$^{-3}$ \M.  Late-time (200 to 700d) optical
spectroscopy and MIR photometry \citep[after 300d, ][]{07od} show
clear evidence of dust formation.  The Ni mass derived including the
MIR black-body emission due to dust indicates a much larger
M(Ni)$\sim2\times10^{-2}$ \M.

The early spectra show the presence of Si II, a boxy \Ha\/
profile, and extreme ($\sim25000$ \kms\/) HV features of the Balmer
lines. The material ejected at high velocity is compatible with a CSM
mass of the order of $\sim10^{-3}$ -- $10^{-4}$ \M\/ located close to
the exploding star.  The interaction of the ejecta with a thin,
moderately dense CSM might increase the luminosity of the light
curves 
and reduce the duration of the plateau (part of the H envelope of the
star is lost and the H mass that recombines is smaller).  Finally, the
expansion velocity and black-body temperature are comparable to those
of type II such as SN 1992H and SN 2004et until the tail.

Late spectra show weaker [O I] than other SNe II, evidence of
ejecta-CSM interaction, and early dust formation.  The boxy line
profile of \Ha\/ was interpreted as the interaction of
spherically symmetric SN ejecta expanding in a medium of low average
density, but with dense clumps.  A similar configuration was invoked
for SN 1988Z \citep {88z}, and allows for the early formation of dust
in a CDS inside clumps that eventually become incorporated within the
SN ejecta.  Radiative transfer models \citep[][]{07od} have provided
estimates of the total dust mass up to $4.2 \times 10^{-3}$ \M, which
may represent only a lower limit due to its clumpy distribution. \\
Only at epochs later than 500d, when the light curve flattens, the
interaction might dominate over other sources of energy.

\od\/ exploded in a LSB galaxy, presumably in a low metallicity
environment with log(O/H) $\sim-4$. In general, low metallicity stars
suffer less mass loss than higher metallicity counterparts, so that it
appears likely that the significant mass loss probed by the
interactions occurred because the progenitor was a super-AGB star that
still preserved H before the final explosion \citep[][]{pumo}.  This
is compatible with the results of our modeling of the SN observables.

\od\/ shares many properties with both type IIn and typical type IIP
objects.  Our analysis has provided indications that: 1) HV features
are present also in CC-SNe; 2) early ejecta-CSM interaction can be
significant also for SNe IIP; 3) dust formation can occur very
early in the SN evolution and affect both the
photometric and the spectroscopic behavior significantly; and 4) extended sampling of
the SED is essential to describe all phenomena characterizing the
CC-SN evolution.  The properties of the "beasts" of the CC-SN zoo do
not cease to be amazing.

\section*{Acknowledgments}

C.I. is grateful to prof. Lucio Patern\`o for useful advice and
suggestions in the earliest stages of his PhD course.  C.I. is also
grateful to Eddie Baron and David Branch for countless discussions and
suggestions, and for hospitality at the University of Oklahoma.  The
authors thank J. Andrews for providing the late time digital spectra
of \od. The author thank also J. Vinko for the useful suggestions.
S.B., F.B., E.C., M.D.V., and M.T. are partially supported by
the PRIN-INAF 2009 "Supernovae Variety and Nucleosynthesis Yields"
and by the grant ASI-INAF I/009/10/0.
M.L.P. acknowledges the financial support by the Bonino-Pulejo
Foundation. The TriGrid VL project, the {\it ``consorzio COMETA''} and
the INAF - Padua Astronomical Observatory are also acknowledged for
computer facilities.  We thank the support astronomers at the
Telescopio Nazionale Galileo, the Copernico Telescope at Cima Ekar,
the 2.2m Telescope at Calar Alto, the Nordic Optical Telescope, the
New Tecnology Telescope, the Galileo Galilei Telescope on Cima Pennar
and the Hale Telescope on Palomar Observatory for performing the
follow-up observations of \od.  This research has made use of the
NASA/IPAC Extragalactic Database (NED) which is operated by the Jet
Propulsion Laboratory, California Institute of Technology, under
contract with the National Aeronautics and Space Administration. This
research has made use of data obtained from the High Energy
Astrophysics Science Archive Research Center (HEASARC), provided by
NASAÕs Goddard Space Flight Center.

\label{lastpage}


\begin{thebibliography}{99}
\bibitem[\protect\citeauthoryear{Andrews et al.}{2010}]{07od} Andrews J.~E., et al., 2010, ApJ, 715, 541
\bibitem[\protect\citeauthoryear{Andrews et 
al.}{2011}]{07it} Andrews J.~E., et al., 2011, ApJ, 731, 47 
\bibitem[\protect\citeauthoryear{Arnett et al.}{1989}]{87a} Arnett W.~D., Bahcall J.~N., Kirshner R.~P., Woosley S.~E., 1989, ARA\&A, 27, 629
\bibitem[\protect\citeauthoryear{Arnett}{1996}]{87a1} Arnett W. D., 1996, Supernovae and Nucleosynthesis, Princeton University Press, Princeton
\bibitem[\protect\citeauthoryear{Balinskaia, Bychkov, 
\& Neizvestnyi}{1980}]{ba80} Balinskaia I.~S., Bychkov K.~V., Neizvestnyi S.~I., 1980, A\&A, 85, L19
\bibitem[\protect\citeauthoryear{Barbon, Ciatti, \& Rosino}{1979}]{barbon} Barbon R., Ciatti F., Rosino L., 1979, A\&A, 72, 287
\bibitem[\protect\citeauthoryear{Baron et al.}{2000}]{99em2} Baron E., et al., 2000, ApJ, 545, 444
\bibitem[\protect\citeauthoryear{Botticella et al.}{2009}]{08s} Botticella M.~T., et al., 2009, MNRAS, 398, 1041
\bibitem[\protect\citeauthoryear{Blondin \& Calkins}{2007}]{c2} Blondin S., Calkins M., 2007, CBET, 1119, 1
\bibitem[\protect\citeauthoryear{Branch et al.}{1981}]{branch} Branch D., Falk S.~W., Uomoto A.~K., Wills B.~J., McCall M.~L., Rybski P., 1981, ApJ, 244, 780
\bibitem[\protect\citeauthoryear{Branch et al.}{2000}]{top} 
Branch D., Jeffery D.~J., Blaylock M., Hatano K., 2000, PASP, 112, 217 
\bibitem[\protect\citeauthoryear{Cappellaro, Danziger, 
\& Turatto}{1995}]{86e} Cappellaro E., Danziger I.~J., Turatto M., 1995, MNRAS, 277, 10
\bibitem[\protect\citeauthoryear{Cappellaro et 
al.}{1997}]{capdecay} Cappellaro E., Mazzali P.~A., Benetti S., Danziger I.~J., Turatto M., della Valle M., Patat F., 1997, A\&A, 328, 203
\bibitem[\protect\citeauthoryear{Cardelli, Clayton, \& Mathis}{1989}]{ca} Cardelli J.~A., Clayton G.~C., Mathis J.~S., 1989, ApJ, 345, 245
\bibitem[\protect\citeauthoryear{Cernuschi, Marsicano, \& Codina}{1967}]{ce} Cernuschi F., Marsicano F., Codina S., 1967, AnAp, 30, 1039
\bibitem[\protect\citeauthoryear{Clocchiatti et al.}{1996}]{92h} Clocchiatti A., et al., 1996, AJ, 111, 1286
\bibitem[\protect\citeauthoryear{Chevalier}{1982}]{chev3} Chevalier R.~A., 1982, ApJ, 258, 790
\bibitem[\protect\citeauthoryear{Chevalier \& Fransson}{1994}]{chev} Chevalier R.~A., Fransson C., 1994, ApJ, 420, 268 
\bibitem[\protect\citeauthoryear{Chevalier}{2008}]{chev2} Chevalier R.~A., 2008, IAUS, 255, 175
\bibitem[\protect\citeauthoryear{Chugai \& Danziger}{1994}]{88z} Chugai N.~N., Danziger I.~J., 1994, MNRAS, 268, 173
\bibitem[\protect\citeauthoryear{Chugai, Chevalier, 
\& Utrobin}{2007}]{chugai} Chugai N.~N., Chevalier R.~A., Utrobin V.~P., 2007, ApJ, 662, 1136
\bibitem[\protect\citeauthoryear{Crockett et al.}{2008}]{cro08} Crockett, R. M. et al., 2008, ApJ, 672, L99
\bibitem[\protect\citeauthoryear{de Blok \& van der Hulst}{1998}]{bev} de Blok W.~J.~G., van der Hulst J.~M., 1998, A\&A, 335, 421
\bibitem[\protect\citeauthoryear{Dessart \& Hillier}{2005}]{99em3} Dessart L., Hillier D.~J., 2005, A\&A, 437, 667
\bibitem[\protect\citeauthoryear{Dwek, Galliano, \& Jones}{2007}]{dw} Dwek E., Galliano F., Jones A.~P., 2007, ApJ, 662, 927
\bibitem[\protect\citeauthoryear{Elmhamdi et al.}{2003}]{99em} Elmhamdi A., et al., 2003, MNRAS, 338, 939 
\bibitem[\protect\citeauthoryear{Elmhamdi, Chugai, \& Danziger}{2003}]{elm} Elmhamdi A., Chugai N.~N., Danziger I.~J., 2003, A\&A, 404, 1077
\bibitem[\protect\citeauthoryear{Ercolano, Barlow, \& Sugerman}{2007}]{87a3} Ercolano B., Barlow M.~J., Sugerman B.~E.~K., 2007, MNRAS, 375, 753
\bibitem[\protect\citeauthoryear{Fassia et al.}{1998}]{fassia} Fassia A., Meikle W.~P.~S., Geballe T.~R., Walton N.~A., Pollacco D.~L., Rutten R.~G.~M., Tinney C., 1998, MNRAS, 299, 150
\bibitem[\protect\citeauthoryear{Fesen et al.}{1999}]{fesen99} Fesen R.~A., et al., 1999, AJ, 117, 725 
\bibitem[\protect\citeauthoryear{Fox et al.}{2009}]{05ip} Fox O., et al., 2009, ApJ, 691, 650
\bibitem[\protect\citeauthoryear{Fransson \& Chevalier}{1987}]{frans} Fransson C., Chevalier R.~A., 1987, ApJ, 322, L15
\bibitem[\protect\citeauthoryear{Fransson \& Chevalier}{1989}]{frans1} Fransson C., Chevalier R.~A., 1989, ApJ, 343, 323
\bibitem[\protect\citeauthoryear{Fryer}{1999}]{fryer} Fryer 
C.~L., 1999, ApJ, 522, 41
\bibitem[\protect\citeauthoryear{Gerardy et al.}{2000}]{98s2} Gerardy C.~L., Fesen R.~A., H{\"o}flich P., Wheeler J.~C., 2000, AJ, 119, 2968
\bibitem[\protect\citeauthoryear{Gerardy et al.}{2002}]{95n} Gerardy C.~L., et al., 2002, ApJ, 575, 1007
\bibitem[\protect\citeauthoryear{Hamuy}{2003}]{hamuy} Hamuy M., 2003, ApJ, 582, 905
\bibitem[\protect\citeauthoryear{Harutyunyan et al.}{2008}]{avik} Harutyunyan A.~H., et al.,  2008, A\&A, 488, 383
\bibitem[\protect\citeauthoryear{Heger et al.}{2003}]{heger} Heger A., Fryer C.~L., Woosley S.~E., Langer N., Hartmann D.~H., 2003, ApJ, 591, 288
\bibitem[\protect\citeauthoryear{Hoyle \& Wickramasinghe}{1970}]{hw} Hoyle F., Wickramasinghe N.~C., 1970, Natur, 226, 62
\bibitem[\protect\citeauthoryear{Immler \& Brown}{2007}]{a1} Immler S., Brown P.~J., 2007, ATel, 1259, 1
\bibitem[\protect\citeauthoryear{Inserra, Baron et al.}{in preparation}]{norman} Inserra C., Baron E., et al, in preparation
\bibitem[\protect\citeauthoryear{Jeffery 
\& Branch}{1990}]{jeffery} Jeffery D.~J., Branch D., 1990, sjws.conf, 149
\bibitem[\protect\citeauthoryear{Kasen \& Woosley}{2009}]{kasen} Kasen D., Woosley S.~E., 2009, ApJ, 703, 2205
\bibitem[\protect\citeauthoryear{Kotak et al.}{2005}]{04dj2} 
Kotak R., Meikle P., van Dyk S.~D., H{\"o}flich P.~A., Mattila S., 2005, 
ApJ, 628, L123
\bibitem[\protect\citeauthoryear{Kotak et al.}{2009}]{04et2} Kotak R., et al., 2009, ApJ, 704, 306
\bibitem[\protect\citeauthoryear{Kozasa, Hasegawa, 
\& Nomoto}{1991}]{kozasa} Kozasa T., Hasegawa H., Nomoto K., 1991, A\&A, 249, 474 
\bibitem[\protect\citeauthoryear{Landolt}{1992}]{landolt} Landolt A.~U., 1992, AJ, 104, 340
\bibitem[\protect\citeauthoryear{Li et al.}{2006}]{li} Li W., Jha S., Filippenko A.~V., Bloom J.~S., Pooley D., Foley R.~J., Perley D.~A., 2006, PASP, 118, 37
\bibitem[\protect\citeauthoryear{Lucy et al.}{1989}]{lucy} Lucy L.~B., Danziger I.~J., Gouiffes C., Bouchet P., 1989, LNP, 350, 164
\bibitem[\protect\citeauthoryear{Lucy et al.}{1991}]{87a4} Lucy E. et al.,1991, in Supernovae, ed. S.E. Woosley, Springer-Verlag, New York, p.82
\bibitem[\protect\citeauthoryear{Maguire et al.}{2010}]{04et} Maguire K., et al., 2010, MNRAS, 284
\bibitem[\protect\citeauthoryear{Mattila et al.}{2008}]{06jc} Mattila S., et al., 2008, MNRAS, 389,141
\bibitem[\protect\citeauthoryear{McGaugh}{1994}]{mcgaugh} McGaugh S.~S., 1994, ApJ, 426,135
\bibitem[\protect\citeauthoryear{Meikle et al.}{2007}]{meikle} Meikle W.~P.~S., et al., 2007, ApJ, 665, 608
\bibitem[\protect\citeauthoryear{Meikle et al.}{2011}]{04dj3} 
Meikle P., et al., 2011, arXiv, arXiv:1103.2885
\bibitem[\protect\citeauthoryear{Mikuz \& Maticic}{2007}]{c1} Mikuz H., Maticic S., 2007, CBET, 1116, 1
\bibitem[\protect\citeauthoryear{Mould et al.}{2000}]{mould} Mould J.~R., et al., 2000, ApJ, 529, 786
\bibitem[\protect\citeauthoryear{Moriya et al.}{2010}]{moriya} Moriya T., Tominaga N., Blinnikov S.~I., Baklanov P.~V., Sorokina E.~I., 2010, arXiv, arXiv:1009.5799
\bibitem[\protect\citeauthoryear{Nozawa et al.}{2003}]{no} Nozawa T., Kozasa T., Umeda H., Maeda K., Nomoto K., 2003, ApJ, 598, 785
\bibitem[\protect\citeauthoryear{Pastorello}{2003}]{pphdt} Pastorello A., 2003, PhD. thesis, Univ.Padova
\bibitem[\protect\citeauthoryear{Pastorello et al.}{2004}]{pa1} Pastorello A., et al., 2004, MNRAS, 347, 74
\bibitem[\protect\citeauthoryear{Pastorello et al.}{2006}]{05cs} Pastorello A., et al., 2006, MNRAS, 370, 1752
\bibitem[\protect\citeauthoryear{Pastorello et al.}{2009}]{05cs2} Pastorello A., et al., 2009, MNRAS, 394, 2266
\bibitem[\protect\citeauthoryear{Patat et al.}{1994}]{p1} Patat F., Barbon R., Cappellaro E., Turatto M., 1994, A\&A, 282, 731
\bibitem[\protect\citeauthoryear{Poole et al.}{2008}]{swift} Poole T.~S., et al., 2008, MNRAS, 383, 627
\bibitem[\protect\citeauthoryear{Pooley et al.}{2002}]{pooley} Pooley D., et al., 2002, ApJ, 572, 932 
\bibitem[\protect\citeauthoryear{Pozzo et al.}{2004}]{98s} Pozzo M., Meikle W.~P.~S., Fassia A., Geballe T., Lundqvist P., Chugai N.~N., Sollerman J., 2004, MNRAS, 352, 457
\bibitem[\protect\citeauthoryear{Pozzo et al.}{2006}]{02hh} Pozzo M., et al., 2006, MNRAS, 368, 1169
\bibitem[\protect\citeauthoryear{Richardson et al.}{2002}]{richardson} Richardson D., Branch D., Casebeer D., Millard J., Thomas R.~C., Baron E., 2002, AJ, 123, 745
\bibitem[\protect\citeauthoryear{Pumo et al.}{2009}]{pumo} Pumo M.~L., et al., 2009, ApJ, 705, L138 
\bibitem[\protect\citeauthoryear{Pumo, Zampieri, \& Turatto}{2010}]{pumo2} Pumo M.~L., Zampieri L., Turatto M., 2010, MSAIS, 14, 123 
\bibitem[\protect\citeauthoryear{Schmidt et al.}{1993}]{90e} Schmidt B.~P., et al., 1993, AJ, 105, 2236
\bibitem[\protect\citeauthoryear{Schlegel}{1990}]{IIn} Schlegel E.~M., 1990, MNRAS, 244, 269 
\bibitem[\protect\citeauthoryear{Schlegel, Finkbeiner, \& Davis}{1998}]{ext} Schlegel D.~J., Finkbeiner D.~P., Davis M., 1998, ApJ, 500, 525
\bibitem[\protect\citeauthoryear{Smartt et al.}{2009}]{smart} Smartt S.~J., Eldridge J.~J., Crockett R.~M., Maund J.~R., 2009, MNRAS, 395, 1409
\bibitem[\protect\citeauthoryear{Smith et al.}{2011}]{smith} 
Smith N., Li W., Filippenko A.~V., Chornock R., 2011, MNRAS, 412, 1522
\bibitem[\protect\citeauthoryear{Smoker, Axon, \& Davies}{1999}]{smo1} Smoker J.~V., Axon D.~J., Davies R.~D., 1999, A\&A, 341, 725
\bibitem[\protect\citeauthoryear{Smoker et al.}{2000}]{smo} Smoker J.~V., Davies R.~D., Axon D.~J., Hummel E., 2000, A\&A, 361, 19
\bibitem[\protect\citeauthoryear{Spyromilio \& Leibundgut}{1996}]{95ad} Spyromilio J., Leibundgut B., 1996, MNRAS, 283, 89
\bibitem[\protect\citeauthoryear{Spyromilio, Leibundgut, \& Gilmozzi}{2001}]{98dl} Spyromilio J., Leibundgut B., Gilmozzi R., 2001, A\&A, 376, 188
\bibitem[\protect\citeauthoryear{Suntzeff \& Bouchet}{1990}]{87a2} Suntzeff N.~B., Bouchet P., 1990, AJ, 99, 650
\bibitem[\protect\citeauthoryear{Szalai et 
al.}{2011}]{04dj} Szalai T., Vink{\'o} J., Balog Z., G{\'a}sp{\'a}r A., Block M., Kiss L.~L., 2011, A\&A, 527, A61
\bibitem[\protect\citeauthoryear{Todini \& Ferrara}{2001}]{tf} Todini P., Ferrara A., 2001, MNRAS, 325, 726
\bibitem[\protect\citeauthoryear{Turatto et al.}{1990}]{t1} Turatto M., Cappellaro E., Barbon R., della Valle M., Ortolani S., Rosino L., 1990, AJ, 100, 771
\bibitem[\protect\citeauthoryear{Turatto, Benetti, \& Cappellaro}{2003}]{t2} Turatto M., Benetti S., Cappellaro E., 2003, fthp.conf, 200
\bibitem[\protect\citeauthoryear{Tully}{1988}]{tully} Tully R. B., 1988, Nearby Galaxies Catalogue, Cambridge University Press, Cambridge and New York, p.221
\bibitem[\protect\citeauthoryear{Zampieri}{2002}]{zamp} Zampieri, L. 2002,  Recent developments in general relativity. 14th SIGRAV Conference on General Relativity and Gravitational Physics, edited by R. Cianci, R. Collina, M. Francaviglia, P. Fr\`e. Milano: Springer.
\bibitem[\protect\citeauthoryear{Zampieri et al.}{2003}]{zamp2} Zampieri L., Pastorello A., Turatto M., Cappellaro E., Benetti S., Altavilla G., Mazzali P., Hamuy M., 2003, MNRAS, 338, 711
\bibitem[\protect\citeauthoryear{Zampieri}{2007}]{zamp3} 
Zampieri L., 2007, AIPC, 924, 358
\bibitem[\protect\citeauthoryear{Utrobin}{2004}]{utrobin} 
Utrobin V.~P., 2004, AstL, 30, 293  
\bibitem[\protect\citeauthoryear{Welch et al.}{2007}]{welch} Welch, D.~L., Clayton, 
G.~C., Campbell, A., Barlow, M.~J., Sugerman, B.~E.~K., Meixner, M., 
\& Bank, S.~H.~R.\ 2007, ApJ, 669, 525
\bibitem[\protect\citeauthoryear{Woosley, Hartmann, \& Pinto}{1989}]{w2} Woosley S.~E., Hartmann D., Pinto P.~A., 1989, ApJ, 346, 395
\bibitem[\protect\citeauthoryear{Woosley, Heger, \& Weaver}{2002}]{woosley2} Woosley S.~E., Heger A., Weaver T.~A., 2002, RvMP, 74, 1015 
\end{thebibliography}
\end{document}